\documentclass[reqno,12pt]{article}
\usepackage{enumerate}
\usepackage{ifpdf}
\usepackage{ae}
\usepackage[T1]{fontenc}
\usepackage[ansinew]{inputenc}
\usepackage{mathrsfs}
\usepackage{amsmath}
\usepackage{amssymb}
\usepackage{graphicx}
\usepackage{color}
\definecolor{darkblue}{cmyk}{0.9,0.9,0,0}
\definecolor{darkgreen}{rgb}{0,0.55,0}
\usepackage[colorlinks=true,linkcolor=darkblue,citecolor=darkblue,urlcolor=darkblue]{hyperref}
\usepackage{epsfig}

\newcommand{\captionfonts}{\it\small}

\makeatletter
\long\def\@makecaption#1#2{
  \vskip\abovecaptionskip
  \sbox\@tempboxa{{\captionfonts #1: #2}}
  \ifdim \wd\@tempboxa >\hsize
    {\captionfonts #1: #2\par}
  \else
    \hbox to\hsize{\hfil\box\@tempboxa\hfil}
  \fi
  \vskip\belowcaptionskip}
\makeatother

\newcommand{\beq}{\begin{equation}}
\newcommand{\eeq}{\end{equation}}
\newcommand{\beqq}{\begin{equation*}}
\newcommand{\eeqq}{\end{equation*}}
\newcommand\beqa{\begin{eqnarray}}
\newcommand\eeqa{\end{eqnarray}}
\newcommand\beqaa{\begin{eqnarray*}}
\newcommand\eeqaa{\end{eqnarray*}}
\newcommand\bea{\begin{array}}
\newcommand\eea{\end{array}}

\def\XXint#1#2#3{{\setbox0=\hbox{$#1{#2#3}{\int}$}
\vcenter{\hbox{$#2#3$}}\kern-.5\wd0}}

\newcommand{\nn}{\nonumber}

\newcommand{\neqa}{\nonumber\end{eqnarray}}
\newcommand{\la}[1]{\label{#1}}

\newcommand{\Tr}{{\rm Tr}}

\renewcommand{\d}{\partial}

\newcommand{\<}{{\langle}}
\renewcommand{\>}{{\rangle}}

\newcommand{\re}{\relax{\rm I\kern-.18em R}}

\renewcommand{\sp}{p\hspace{-.40em}/}

\def\su2{{SU(2)}}

\def\a{{\alpha}}

\def\[{\left[}
\def\]{\right]}

\def\l{\lambda}

\def\a{\alpha}

\def\({\left(}
\def\){\right)}
\def\[{\left[}
\def\]{\right]}

\def\<{\langle}
\def\>{\rangle}

\def\i2{\frac{i}{2}}

\newcommand{\N}{\mathcal{N}}
\def\O{{\mathcal O}}

\def\spi{\relax{\rm \pi\kern-0.5em /}}
\def\sA{\relax{\rm A\kern-0.5em /}}
\def\sp{\relax{\rm p\kern-0.5em /}}
\def\sd{\relax{\rm \d\kern-0.5em /}}
\def\sk{\relax{\rm k\kern-0.5em /}}
\def\sn{\relax{\rm n\kern-0.5em /}}
\def\sl{\relax{\rm l\kern-0.5em /}}
\def\sP{\relax{\rm P\kern-0.7em /}}
\def\sBethe{\relax{\rm \Bethe\kern-0.5em /}}

\def\bu{{\bf u}}
\def\bbu{{\bf\bar u}}

\usepackage{varioref}
\usepackage{makeidx}
\makeindex

\usepackage[english]{babel}
\begin{document}

\thispagestyle{empty}

\renewcommand{\thefootnote}{\fnsymbol{footnote}}
\setcounter{page}{1}
\setcounter{footnote}{0}
\setcounter{figure}{0}
\begin{center}
$$$$
{\Large\textbf{\mathversion{bold}
On four-point functions and integrability \\ in $ \mathcal{N}=4 $ SYM: from weak to strong coupling}\par}

\vspace{1.0cm}

\textrm{Jo\~ao Caetano$^{a,b,c}$, \ Jorge Escobedo$^{a,b}$}
\\ \vspace{1.2cm}
\footnotesize{

\textit{$^a$Perimeter Institute for Theoretical Physics\\ Waterloo,
Ontario N2L 2Y5, Canada}  \\
\vspace{4mm}
\textit{$^b$Department of Physics and Astronomy \& Guelph-Waterloo Physics Institute,\\
University of Waterloo, Waterloo, Ontario N2L 3G1, Canada} \\
\vspace{4mm}
\textit{$^c$Centro de F\'isica do Porto e Departamento de F\'isica e Astronomia,\\
Faculdade de Ci\^encias da Universidade do Porto,\\
Rua do Campo Alegre, 687, 4169-007 Porto, Portugal} \\
\vspace{5mm}
\small\texttt{jd.caetano.s,jescob@gmail.com} \\
}

\par\vspace{1.5cm}

\textbf{Abstract}\vspace{2mm}
\end{center}

\noindent
Using integrability techniques, we compute four-point functions of single trace gauge-invariant operators in $\N=4$ SYM to leading order at weak coupling. Our results are valid for operators of arbitrary size. In particular, we study the limit in which two of the four operators are taken to be much smaller than the others. We show that in this limit our weak coupling result matches with the strong coupling result in the Frolov-Tseytlin limit. 

\vspace*{\fill}

\setcounter{page}{1}
\renewcommand{\thefootnote}{\arabic{footnote}}
\setcounter{footnote}{0}

\newpage

 \def\nref#1{{(\ref{#1})}}

\tableofcontents

\section{Introduction}
Recently, there has been a lot of activity in computing three-point functions of single trace operators in the context of the AdS/CFT correspondence \cite{adscft1,adscft2,adscft3}, both at weak \cite{Okuyama:2004bd,Roiban:2004va,Alday:2005nd,paper1,paper2} and strong coupling \cite{recentpapers1,recentpapers2,Roiban:2010fe}.  The computation of certain four-point functions at strong coupling has also been performed \cite{Buchbinder:2010ek}-\cite{Arutyunov:2003ae}. In this paper we use the integrability-based approach to correlation functions introduced in \cite{paper1} to compute four-point functions of single trace gauge-invariant operators in the $SU(2)$ sector of $\N=4$ SYM. Furthermore, motivated by the weak/strong coupling match for three-point functions in the classical limit reported in \cite{paper2}, we will show that the same exact match occurs for the four-point function of two heavy non-BPS operators and two light BPS operators. In this introduction, we anticipate some of our main results and set our notations and conventions.

Conformal invariance of $\N=4$ SYM fixes the two-point and three-point functions of its operators to take the  form
\beqq\la{2pt3pt}
G_2(x_1,x_2) \equiv \< \O_i(x_1) \bar \O_i(x_2)\> =\frac{1}{|x_{12}|^{2\Delta_i}} \, ,
\eeqq
and
\beqq
G_3(x_1,x_2,x_3) \equiv \<\O_1 (x_1) \O_2 (x_2) \O_3 (x_3) \> = \frac{1}{N} \frac{C_{123}}{|x_{12}|^{\Delta_1+\Delta_2-\Delta_3} |x_{23}|^{\Delta_2+\Delta_3-\Delta_1} |x_{31}|^{\Delta_3+\Delta_1-\Delta_2}} \, ,
\eeqq
where $x_{ij}\equiv x_i-x_j$, $\Delta_i$ are the dimensions of the operators, $C_{123}$ are the so-called structure constants and we have normalized the two-point functions to one. The structure constants are not completely unambiguous, since multiplying $\O_i$ by a phase will change $C_{123}$ by that phase. On the other hand the absolute value $|C_{123}|$ is unambiguously defined. As opposed to the two- and three-point functions, the form of four-point functions is not uniquely determined in $\N=4$ SYM. Indeed, conformal invariance only tells us that the four-point function will depend on the cross-ratios 
\beqq
 a\equiv\frac{x^{2}_{12} x^{2}_{34}}{x^{2}_{13} x^{2}_{24}} \qquad , \qquad  b\equiv\frac{x^{2}_{12} x^{2}_{34}}{x^{2}_{14} x^{2}_{23}}
\eeqq
that we can form with the positions of the operators. Namely, in the planar limit
\beq
G_4(x_1,x_2,x_3,x_4) \equiv \<\O_1 (x_1) \O_2 (x_2) \O_3 (x_3) \O_4 (x_4) \>_{\text{connected}}= \frac{1}{N^2} f\(a,b\) \prod_{i<j}^4 |x_{ij}|^{\Delta/3 - \Delta_i - \Delta_j} \, , \la{4pt}
\eeq
where $\Delta=\sum_{i=1}^4 \Delta_i$. 

At weak coupling, the tree-level computation of correlation functions is simply given by summing over all possible Wick contractions between the (constituents of the) single trace operators.  Each single trace operator is an eigenvector of the one-loop dilatation operator. This is necessary to lift the degeneracy present at 't Hooft coupling $\l=0$, where there is a large number of operators with the same classical dimension \cite{Beisert:2002bb}. The problem of performing the Wick contractions between these eigenvectors is purely combinatorial and can be tackled by exploting the integrability of the theory using the tools developed in \cite{paper1}. 

We will consider single trace operators $\O_i$ made out of $L_i$ scalars. Then, at tree level, 
\beqq
G_2(x_1,x_2)= \frac{1}{|x_{12}|^{2L_i}} \qquad , \qquad
G_3(x_1,x_2,x_3)= \frac{1}{N} \frac{C_{123}}{\prod\limits_{i<j}|x_{ij}|^{2l_{ij}}}
\eeqq
where $l_{ij}$ is the number of Wick contractions between operators $\O_i$ and $\O_j$. Note that given the lengths of the three operators the number of Wick contractions between the different single traces is uniquely fixed for three-point functions ($2\,l_{12}=L_1+L_{2}-L_{3}$, etc). This is not the case for four-point functions. Instead\footnote{For non-planar diagrams, each $C_{1234;\{l_{ij}\}}$ will come with additional $1/N$ factors. However, in this paper we will focus on planar diagrams, such that each $C_{1234;\{l_{ij}\}}$ is simply a number.}
\beq
G_4(x_1,x_2,x_3,x_4) = \frac{1}{N^2} \sum_{\text{all possible $\{l_{ij}\}$}} \frac{C_{1234;\{l_{ij}\}}}{\prod\limits_{i<j} |x_{ij}|^{2l_{ij}}} \la{sum} \, .
\eeq
As explained in this paper, each $C_{1234;\{l_{ij}\}}$ can be computed using the integrability techniques developed in \cite{paper1}. In other words, we can fix the function $f(a,b)$ in (\ref{4pt}) to leading order at weak coupling using integrability.

\begin{figure}[t]
\centering
\def\svgwidth{15cm}
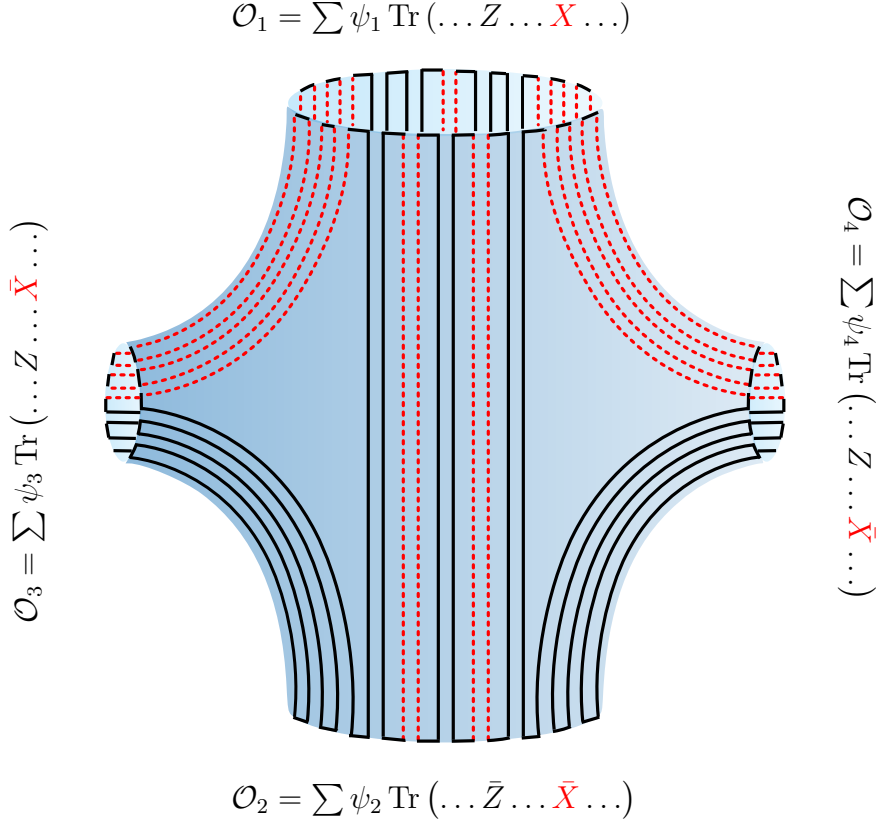
\caption{Four-point function of $SU(2)$ operators at tree level. All contractions are such that R-charge is preserved. The black (solid) lines represent vacuum fields, while the red (dashed) lines represent excitations. Our setup is such that operators $\O_3$ and $\O_4$ do not interact among themselves. This restriction can be trivially relaxed at weak coupling using integrability techniques, see section \ref{anothersetup}.}
\la{generalsetup}
\end{figure}
We can consider four operators $\O_i$ such that only one choice of $\{l_{ij}\}$ survives due to R-charge conservation. This is the case for the setup considered in figure \ref{generalsetup}, for which our convention for the charges of the operators is the following
\beq\la{Vacchoicestrong}
\begin{array}{cccll}
& \text{vacuum} & \text{excitations} && \text{notations}\\ \hline
\O_1 & Z & X && \#\{X, Z\}= \{J_1,J_2\} \\
\O_2 & \bar Z & \bar  X && \#\{\bar X,\bar Z\}= \{J_1-j_1-k_1,J_2+j_2+k_2\}  \\
\O_3 & Z & \bar X && \#\{\bar X,Z\}= \{j_1,j_2\} \\
\O_4 & Z & \bar X && \#\{\bar X,Z\}= \{k_1,k_2\} 
\end{array}
\eeq
while the total lengths of the four operators will be denoted by
\beq
L_1 \equiv J_1+J_2 \, , \qquad L_2 \equiv J_1+J_2 -j_1+j_2-k_1+k_2 \, , \qquad L_3 \equiv j_1+j_2 \, , \qquad L_4 \equiv k_1+k_2 \, . 
\la{lengthsops}
\eeq
For example $l_{23}=j_2$, $l_{24}=k_2$, etc. More importantly, $l_{34}=0$. We will demonstrate analytically and numerically that in the classical limit  $L_1, L_2\gg L_3,L_4$ with $\O_1 \simeq \bar \O_2$, the four-point function factorizes as
\beq
C_{1234,\{l_{ij}\}} \simeq C_{123} C_{124}   \, ,
\la{fac1}
\eeq
which is somehow natural given figure \ref{generalsetup}. 

We will also consider more general four-point functions, for which different $\{l_{ij}\}$ are allowed. In this case, we will explain that in the classical limit the term with $l_{34}=0$ in (\ref{sum}) dominates over the terms with $l_{34}\neq 0$.\footnote{We do predict the form of these other terms in the coherent state language. It would be very interesting to refine the strong coupling computation in order to probe these terms and verify if a match between strong and weak coupling holds for them as well.} Furthermore for these more general four-point functions, we will show that in the classical limit
\beq
C_{1234,\{l_{ij}\}} \simeq C_{123} C_{124}  \,,\quad \text{for $l_{34}=0$} \la{fac2} \, .
\eeq
The factorization (\ref{fac1}) and (\ref{fac2}) have interesting implications. A similar factorization of the four-point function of two large operators and two small operators was reported in \cite{Buchbinder:2010ek} at strong coupling. In the Frolov-Tseytlin limit a match between weak and strong coupling for three-point functions was recently demonstrated in \cite{paper2}. Hence, the factorization presented here extends this match to four-point functions of two large operators and two small operators. The numerical check alluded to above is extremely important in validating this match since we know that the issue of \textit{back-reaction} is very important \cite{paper2}. It could have been that further subtle issues arise for four-point functions. Our numerics indicate that this is not the case.

The paper is structured as follows: in section \ref{sectionstrong} we review the holographic computation of the three- and four-point functions relevant to this paper. In section \ref{sectionweak} we use integrability techniques to compute tree-level four-point functions at weak coupling. In section \ref{sectionmatch} we show numerically and analytically that in the Frolov-Tseytlin limit, the computations at weak and strong coupling of the four-point function of two heavy and two light operators match exactly. In section \ref{anothersetup} we study a more general class of four-point function than the one considered in section  \ref{sectionstrong} and study their classical limit using coherent states.  Appendix \ref{tailoringtools} contains a short review of the main integrability tools to compute correlation functions in $\N=4$ SYM, while in appendix \ref{genint} we use integrability techniques to obtain a general formula for the four-point functions considered in section \ref{anothersetup}. Finally, appendix \ref{appcodes} contains the Mathematica codes needed to compute the main results of this paper.


\section{Strong coupling} \la{sectionstrong}

\begin{figure}[t]
\centering
\def\svgwidth{11.5cm}
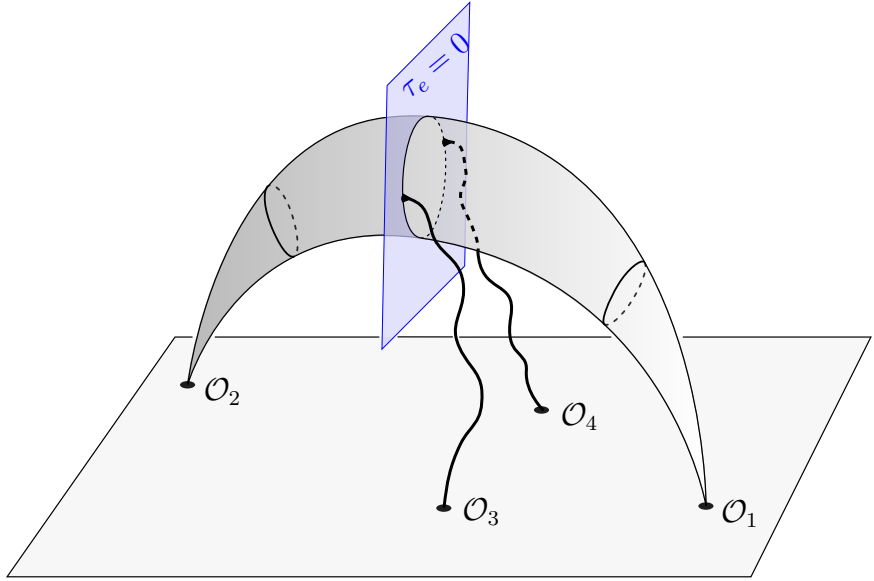
\caption{The strong coupling computation of the four-point function of two heavy and two light operators factorizes into a product of two three-point functions, each of which involves a two-dimensional integration of a boundary-to-bulk propagator over the full string worldsheet. In the Frolov-Tseytlin limit both integrations become localized in the slice
$\tau_e=0$ and only the integrals over $\sigma$ survive. In this way
the computations at weak and strong coupling are matched.}
\la{4ptstrongcoupling}
\end{figure}
In \cite{Buchbinder:2010ek}, the holographic computation of four-point functions $G_4(x_1,x_2,x_3,x_4)$ of single trace gauge-invariant operators was considered. It was argued that if the charges of the operators $\mathcal{O}_1$ and $\mathcal{O}_2$ are much larger than those of the operators $\mathcal{O}_3$ and $\mathcal{O}_4$, such that $\O_1 \simeq \bar\O_2$, then the four-point function factorizes into a product of two three-point functions
\beq
G_4 (x_1,x_2,x_3,x_4) = \frac{G_3(x_1,x_2,x_3) \, G_3(x_1,x_2,x_4)}{G_2(x_1,x_2)} \, .
\la{factor}
\eeq
This is depicted in figure \ref{4ptstrongcoupling}. If we take $\O_3$ and $\O_4$ to be light chiral primaries and consider the setup shown in \eqref{Vacchoicestrong}, then only one choice of $\{l_{ij}\}$ survives and expression (\ref{factor}) simply translates into
\begin{equation}\label{factor2}
C^{\bullet \bullet \circ \circ}_{1234} = C^{\bullet \bullet \circ}_{123} C^{\bullet \bullet \circ}_{124} \, ,
\end{equation}
where $\bullet$ denotes a non-BPS operator, while $\circ$ denotes a BPS operator. Consequently, we just need to know how to compute the holographic three-point function of two large operators and a light BPS operator. 

In order to compare the strong coupling results with the classical limit of our weak coupling computation, we need to take the Frolov-Tseytlin limit \cite{FT} of \eqref{factor2}.\footnote{Recall that given a classical string solution with total charge $J$, the Frolov-Tseytlin limit corresponds to taking $\lambda, J \to \infty$ with $\lambda/J^2 \ll 1$. For more details, see section 2 of \cite{paper2}.} Luckily, we can simply use the results of \cite{paper2} to compute the three-point functions appearing on the r.h.s of the factorization formula \eqref{factor2}. Using the charges of the four operators as given in \eqref{Vacchoicestrong}, at the end of the day one obtains (see section 2 of  \cite{paper2} for details)
\beq
r_{123} \equiv \left| \frac{C^{\bullet\bullet\circ}_{123}}{C^{\circ\circ\circ}_{123}} \right| = \left|  \frac{1}{{v_1^{j_1}  \, \bar v_2^{j_2}}} \int\limits_0^{2\pi} \frac{d\sigma}{2\pi} \,  u_1^{j_1} \,  \bar u_2^{j_2} \right|_{\tau_e=0}\, , 
\la{ratio123}
\eeq
\beq
r_{124} \equiv \left| \frac{C^{\bullet\bullet\circ}_{124}}{C^{\circ\circ\circ}_{124}} \right| = \left|  \frac{1}{{v_1^{k_1}  \, \bar v_2^{k_2}}} \int\limits_0^{2\pi} \frac{d\sigma}{2\pi} \,  u_1^{k_1} \,  \bar u_2^{k_2} \right|_{\tau_e=0}\, ,
\la{ratio124}
\eeq
where
\beq
C_{123}^{\circ\circ\circ}=
L_1 \, v_1^{j_1} \bar v_2^{j_2}\sqrt{\frac{j_1!\, j_2!}{(j_1+j_2-1)!}} \, , \qquad v_a=\sqrt{\frac{J_a}{L_1}} \, ,
\la{bpssimp}
\eeq
$C^{\circ\circ\circ}_{124}$ is of course exactly the same as $C^{\circ\circ\circ}_{123}$ with $j_i \to k_i$ and $u_1$, $\bar u_2$ are determined by the classical string solution in $S^5$ dual to operators $\O_1$ and $\O_2$. Recall that one takes the ratio of the structure constant of interest to the three-point function of three BPS operators in order to remove any dependence on the normalization convention. We also take the absolute value of the ratio to avoid the ambiguity that arises when multiplying each operator in the three-point function by a phase. Hence, the factorization formula \eqref{factor2} in the Frolov-Tseytlin limit reads
\beq 
r_{1234} \equiv \left| \frac{C_{1234}^{\bullet \bullet \circ \circ}}{C_{123}^{\circ\circ\circ}C_{124}^{\circ\circ\circ}} \right| = r_{123} \, r_{124} \, .
\la{ratio1234}
\eeq
This is the main formula of this section. 

A few remarks are in order. The charges of the operators in the four-point function in the factorization formula \eqref{factor}, or equivalently in \eqref{ratio1234}, are those specified in \eqref{Vacchoicestrong}. However, the charges of the operators entering the three-point functions in \eqref{factor} and \eqref{ratio1234} must be, due to charge conservation, \textit{slightly} different. This point was already raised in \cite{Buchbinder:2010ek}; however, the operators considered therein were made out of the complex scalars $Z$ and $\bar Z$ only. Since in this paper we are considering operators that also have $X$ and $\bar X$, let us give the correct prescription for the charges of the operators in the three-point functions entering the factorization formulas above. Given the charges shown in \eqref{Vacchoicestrong}, the charges of the operators in the three-point function $C_{123}$ are $\{J_1,J_2\}$, $\{J_1-j_1,J_2+j_2\}$ and $\{j_1,j_2\}$, respectively. Similarly, the charges of the operators in $C_{124}$ are $\{J_1,J_2\}$, $\{J_1-k_1,J_2+k_2\}$ and $\{k_1,k_2\}$.


\section{Weak coupling} \la{sectionweak}

In this section, we will describe the computation at weak coupling of tree-level four-point functions of generic operators in the $ SU(2) $ sector of $ \mathcal{N}=4 $ SYM. In the spin chain language, each single-trace operator $\mathcal{O}_i$ is represented by a Bethe state on a closed spin chain \cite{MZ,Staudacher} that we denote by $|\Psi_i \rangle$. For example, for an operator made out of $L-N$ complex scalars $Z$ and $N$ complex scalars $X$, we have
\beq
|\Psi \rangle = \sum_{1\le n_1<\dots <n_N \le L} \psi_{n_1, \dots,n_N} \Tr\Big(Z\dots Z \underset{\underset{n_1}{\downarrow}}{X}Z \dots Z\underset{\underset{n_2}{\downarrow}}{X}Z\dots \Big) \, ,
\la{operators}
\eeq
where $\psi_{n_1,\dots,n_N}$ is the so-called Bethe wave function, whose specific form can be found in \cite{paper1}. We can think of the scalars $X$ as excitations propagating in the $Z$ vacuum. These excitations are conveniently parametrized by their rapidities or Bethe roots $u_i = \frac{1}{2} \cot \frac{p_i}{2}$, where $p_i$ are the momenta of the excitations.

\begin{figure}
\centering
\def\svgwidth{10.5cm}
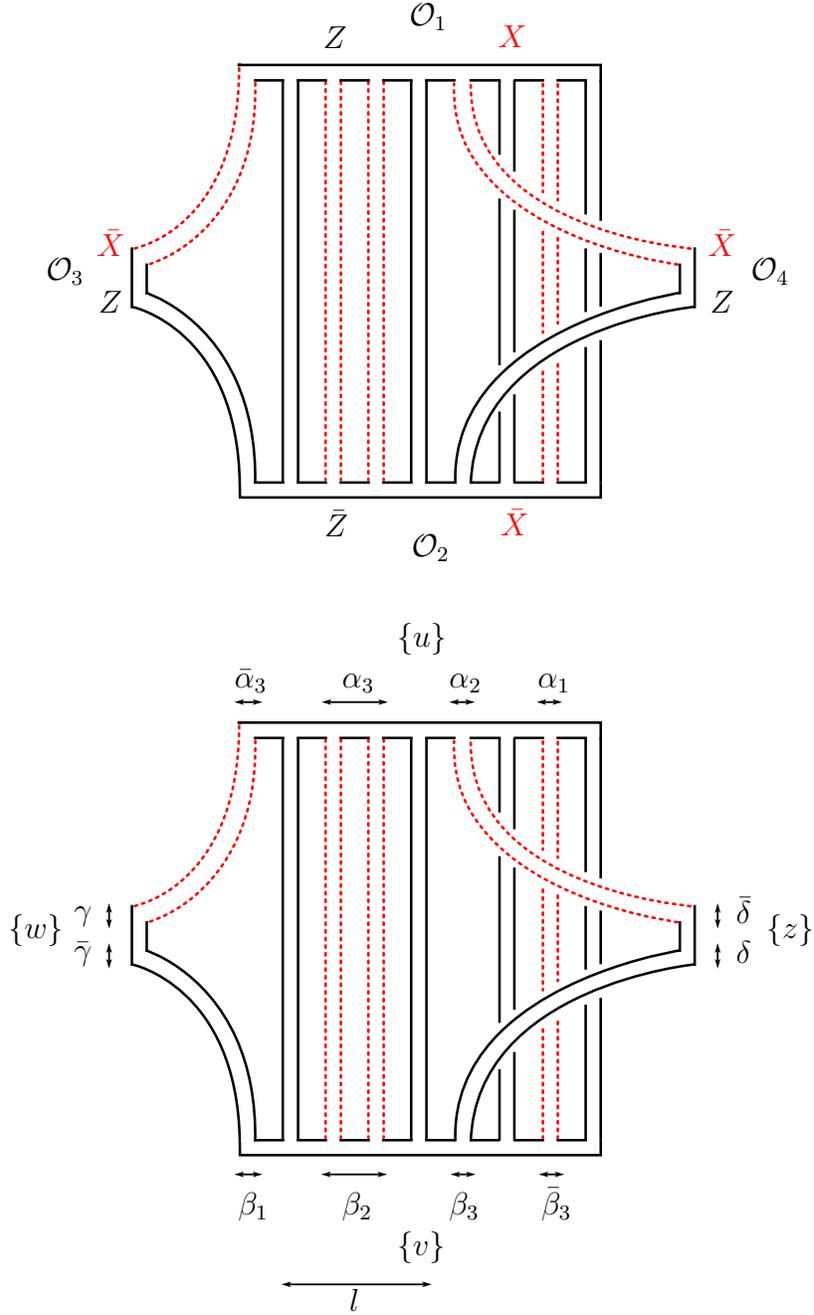
\caption{Setup for the computation of the four-point function at weak coupling. The black (solid) lines represent vacuum fields, while the red (dashed) lines represent excitations. The top figure shows the vacuum ($Z$ or $\bar Z$) and excitation choice ($X$ or $\bar X$) for each operator. The figure at the bottom shows the labelling of the excitations of each operator and the different partitions needed to perform the Wick contractions.}
\la{setuppartitions}
\end{figure}
We will consider the configuration in figure \ref{setuppartitions}. Below we will show how to compute the quantity $C_{1234}$ appearing in \eqref{sum}, where we dropped the subscript $\{l_{ij}\}$ given that in our setup there is only one such choice, which is completely determined by the charges of the operators given again by \eqref{Vacchoicestrong}.

We should stress that in this section, we are \textit{not} assuming that $\O_3$ and $\O_4$ are small operators. Hence, all the results from this section are valid for generic $SU(2)$ operators obeying the setup shown in figure \ref{setuppartitions}. Also, as we mentioned in the introduction, the configuration we are considering does not include interactions between operators $\O_3$ and $\O_4$. However, it is trivial to modify the formulas we present below to consider the more general case in which $\O_3$ and $\O_4$ interact with each other.\footnote{We consider such four-point functions in section \ref{anothersetup}.}

\subsection{Four-point functions by brute force}
We can always do a brute force computation to determine $C_{1234}$ using the explicit form of the Bethe states representing each of the operators, see \eqref{operators}. Then, denoting the lengths of each operator by $L_i$, see \eqref{lengthsops}, we have 
\begin{align}
C_{1234}=&\,\, \Omega \, \sum_{l=0}^{L_1 - j_1-k_1} \sum_{|\a_1| = 0}^{\min{\{l,J_1-j_1-k_1\}}} \sum_{1\leq n_1 < ... < n_{|\a_1|} \leq l} \,\,\, \sum_{1\leq m_{1} < ... < m_{J_1 -j_1 -k_1-|\a_1|} \leq L_1 - j_1 -k_1 -l} \nn \\
&\times \, \psi^{(1)}_{n_1, ..., n_{|\a_1|},l+1,...,l+k_1, k_1+l+m_{1} ,..., k_1 +l+ m_{ J_1 - j_1 - k_1-|\a_1|}, L_1 -j_1+1, ..., L_1} \nn \\
&\times \, \psi^{(2)}_{L_2 - (L_4 - k_1 +l + m_{J_1 - j_1 - k_1- |\a_1|})+1, ...,L_2 - (L_4 - k_1 +l + m_{1})+1, L_2 - n_{|\a_1|}+1, ..., L_2-n_1+1 } \nn \\
&\times \, \psi^{(3)}_{1,2,...,j_1}\, \psi^{(4)}_{L_4 - k_1 +1,...,L_4}
\la{4ptbruteforce}
\end{align}
where $ \psi^{(i)} $ is the Bethe wave function associated to operator $\mathcal{O}_i$, $|\a_1|$ is the number of Bethe roots in partition $\a_1$, see figure \ref{setuppartitions}, and\footnote{We can also include a symmetry factor in $\Omega$ to take into account the case when $\O_3$ or $\O_4$ are dropped from the four-point function:
\beq
\Omega=\sqrt{\frac{L_1 L_2 L_3^{\delta_{L_3>0}} L_4^{\delta_{L_4>0}}}{\N_1 \N_2 \N_3 \N_4}} \, \frac{1}{(L_1-j_1-k_1)\bigl(\Theta(L_3)\Theta(L_4)-\delta_{L_3>0}\delta_{L_4>0}\bigr)+1} \, ,
\eeq
with $\Theta(x) $ being the Heaviside theta function, such that $\Theta(x)=1,$ for $x\geq 0$ and $\Theta(x)=0$, for $x<0$. In this case we recover the brute force formula for three-point functions (see equation (85) of \cite{paper1}).}
\beq
\Omega=\sqrt{\frac{L_1 L_2 L_3 L_4}{\N_1 \N_2 \N_3 \N_4}} \, .
\la{omega}
\eeq  
This factor takes into account the equivalent ways of breaking the spin chains (i.e.\ due to cyclicity we can rotate each chain before cutting it). Finally, $\N_i$ denotes the norm of operator $\O_i$, which is defined as
\beqq
\mathcal{N}_i= \sum_{1\le n_1 < \dots < n_{N_i} \le L_i} \(\psi^{(i)}_{n_1,\dots,n_{N_i}}\)^* \(\psi^{(i)}_{n_1,\dots,n_{N_i}}\) \, ,
\eeqq
where $N_i$ is the number of excitations of operator $\O_i$.

Since each Bethe wave function $\psi^{(i)}$ has $N_i!$ terms, we see that for large $N_i,L_i$ equation \eqref{4ptbruteforce} is computationally extremely inefficient due to the huge number of terms involved. Below, we will see that using integrability techniques, we are able to simplify the computation of $C_{1234}$ significantly.

\subsection{Four-point functions from integrability}
The combinatorial problem associated with the multiple Wick contractions required to compute the tree-level four-point function can be solved using the integrability tools introduced in \cite{paper1}. One simply needs to follow these steps:

\begin{itemize}
\item{We cut each of the closed spin chains 1 and 2 into four open subchains. Formally this means that we write each of the states $ |\Psi_1\rangle  $ and $ |\Psi_2\rangle $ as a linear combination of tensor products of four states in open subchains. Schematically, i.e.\ leaving out the sums over the different partitions and the factors arising from cutting the chains (see appendix \ref{tailoringtools}), and using the notation in figure \ref{setuppartitions}:
\beqq
|\Psi_1\rangle \to | \a_1 \> \otimes | \a_2 \> \otimes | \a_3 \> \otimes | \bar\a_3 \> \, , \qquad |\Psi_2\rangle \to | \beta_1 \> \otimes | \beta_2 \> \otimes | \beta_3 \> \otimes | \bar\beta_3 \> \, . \nn
\eeqq}
\item{Similarly, we cut each of the closed spin chains 3 and 4 into two open subchains, i.e.\ we write each of the states $ |\Psi_3\rangle  $ and $ |\Psi_4\rangle $ as a linear combination of tensor products of two states in open subchains. Schematically
\beqq
|\Psi_3\rangle \to | \gamma \> \otimes | \bar\gamma \> \, , \qquad |\Psi_4\rangle \to | \delta \> \otimes | \bar\delta \>  \, . \nn
\eeqq}
\item{In order to perform the Wick contractions among the operators in the four-point function, we first need to flip some subchain states. Again, leaving out the sum over partitions and the relevant factors coming from the flipping procedure, we schematically have
\beqq
|\Psi_1\rangle \to | \a_1 \> \otimes | \a_2 \> \otimes | \a_3 \> \otimes \< \bar\a_3^* | \, , \qquad |\Psi_2\rangle \to | \beta_1 \> \otimes \< \beta_2^* | \otimes \< \beta_3^* | \otimes \< \bar\beta_3^* | \, . \nn
\eeqq
\beqq
|\Psi_3\rangle \to | \gamma \> \otimes \< \bar\gamma^* | \, , \qquad |\Psi_4\rangle \to | \delta \> \otimes \< \bar\delta^* |  \, . \nn
\eeqq}
\item{We contract, or sew, the different subchain states as shown in figure \ref{setuppartitions}. This involves the computation of scalar products of Bethe states. To efficiently perform them, we use the new recursion relation for $SU(2)$ scalar products derived in \cite{paper1}, which we quickly review in appendix \ref{tailoringtools}.}
\item{Finally, we normalize the states and sum over the distinct ways of breaking them}.
\end{itemize}
All the necessary tools to explicitly perform this computation are reviewed in appendix \ref{tailoringtools}, where all the notation used in the formula below is explained in detail. After all the dust has settled, we get
\begin{align}
C_{1234}=& \,\, \Omega \,
\sum_{l=0}^{L_1 - j_1 - k_1}\sum_{\a_1 \cup \bar{\a}_1 =\{u\}} \sum_{\a_2 \cup \bar{\a}_2 = \bar{\a}_1} \sum_{\a_3 \cup \bar{\a}_3 = \bar\a_2} \sum_{\substack{\beta_2 \cup \bar{\beta}_2 = \{v\} \\ |\beta_2| = |\a_3| \\ |\bar{\beta}_2| = |\a_1 |}} e^{\bar\a_1}_{L_1 -j_1 -k_1-l} \, e^{\bar\a_2}_{k_1} \, \frac{f^{\a_1 \bar\a_1} f^{\a_1 \a_1}_{<} f^{\a_2 \bar\a_2} f^{\a_2 \a_2}_{<}}{f^{\{u\} \{u\}}_{<}} \nn \\
&\times \, e^{\bar\a_3}_{l +j_1+1} \, f^{\a_3 \bar\a_3} f^{\bar\a_3 \bar\a_3}_{>} f^{\a_3 \a_3}_{<} \, e^{\{v\}}_{j_2+l+1} \, e^{\bar{\beta}_2}_{L_2 - j_2-l} \, \frac{f^{\beta_2 \bar{\beta}_2} f^{\beta_2 \beta_2}_{>} f^{\bar{\beta}_2 \bar{\beta}_2}_{>}}{f^{\{v\} \{v\}}_{<}} \,  e^{\{z\}}_{L_4+1} \, \frac{f^{\{z\} \{z\}}_{>}}{f^{\{z\} \{z\}}_{<}}  \nn \\
&\times \, \< \beta_2^{\ast} | \a_3 \> \< \bar{\beta}_3^{\ast}| \a_1 \> \< \{z^*\} | \a_2 \> \<\bar{\a}_3^{\ast} | \{w\} \> \, .
\la{tail}
\end{align}

In section \ref{sectionmatch}, we will use formula \eqref{tail} to compute numerically tree-level four-point functions that satisfy the setup of figure \ref{setuppartitions}. We will then extrapolate our numerical results to the case in which the lengths of $\O_1$ and $\O_2$ go to infinity and see that in this particular limit, the weak coupling result factorizes as in \eqref{fac1} and hence matches the strong coupling result in the Frolov-Tseytlin limit \eqref{factor2} (or equivalently, \eqref{ratio1234}).\footnote{Note that a priori it is not obvious at all that \eqref{tail} factorizes as in \eqref{factor2} when $L_1,L_2 \to \infty$. However, in section \ref{sectionmatch} we provide an analytic proof of the weak/strong coupling match from another perspective, namely, by representing $\O_1$ and $\O_2$ by coherent states.}

Let us stress that the integrability-based formula \eqref{tail} proves to be far more efficient than the brute force computation \eqref{4ptbruteforce}.  For the reader's convenience, we present in appendix \ref{appcodes} the Mathematica codes needed to compute four-point functions using both formulas, as well as some specific examples showing how to use the codes. As noted in that appendix, even when the number of excitations and lengths of the operators are not too large, the brute force formula (\ref{4ptbruteforce}) becomes computationally much slower than formula \eqref{tail}.


\section{Weak/strong coupling match} \la{sectionmatch}
In this section we will show that in the classical limit, the four-point function $C_{1234}^{\bullet \bullet \circ \circ}$ of two large non-BPS operators and two small BPS operators computed at weak coupling matches the strong coupling result. First, we will provide some numerical evidence to support this claim. Namely, we will use the weak coupling result \eqref{tail} to compute the four-point function of interest and then extrapolate numerically our results to the case when the lengths of the two large operators go to infinity. By doing so, we will obtain a numerical match with the strong coupling result evaluated using equation \eqref{ratio1234}. We will then provide an analytic proof of this match by representing the two large operators by coherent states.

\subsection{Numerics: $SU(2)$ folded string}
Let us consider the following operators in the four-point function represented in figure \ref{setuppartitions}. Operators $\O_3$ and $\O_4$ are taken to be the following small BPS operator:
\beq
\O_3 = \O_4 = 2\,\Tr(ZZ\bar X\bar X) + \Tr(Z\bar XZ\bar X) \, .
\eeq
Operators $\O_1$ and $\O_2$ are taken to be large non-BPS operators dual to the folded string with unit mode number. This is very similar to the three-point function setup of two large operators and a light BPS operator considered in \cite{paper2}. We can directly borrow the results from that paper to compute the structure constants appearing on the r.h.s of \eqref{ratio1234} in the Frolov-Tseytlin limit (we refer the reader to section 3 of \cite{paper2} for details). We obtain
\begin{align}
r_{1234}=  \frac{\pi^2 \, q^2 \, (1-q)^2 \, _2F_1 (\frac{1}{2},\frac{5}{2};2;q)^2}{16 \,  \alpha^2 (1-\alpha)^2  \, K(q)^2 } \, ,
\la{intsigma}
\end{align}
where $\alpha$ is the filling fraction of $\O_1$ and $q$ is related to it by
\beq
\alpha\equiv \frac{J_1}{J_1+J_2}=1- \frac{\text{E}(q)}{\text{K}(q)} \la{alpha} \, ,
\eeq
with $\text{E}(q)= \,\, $\verb"EllipticE[q]",  $\text{K}(q)=\,\,$\verb"EllipticK[q]" in Mathematica.
Hence, given a filling fraction $\a$, we can find the corresponding value of $q$ from \eqref{alpha} and plug everything into \eqref{intsigma} to obtain a number, which will be the analytical prediction from strong coupling. The first column in table \ref{tableratios} has different such predictions for three different filling fractions.

\begin{figure}[t]
\centering
\includegraphics[width=11cm]{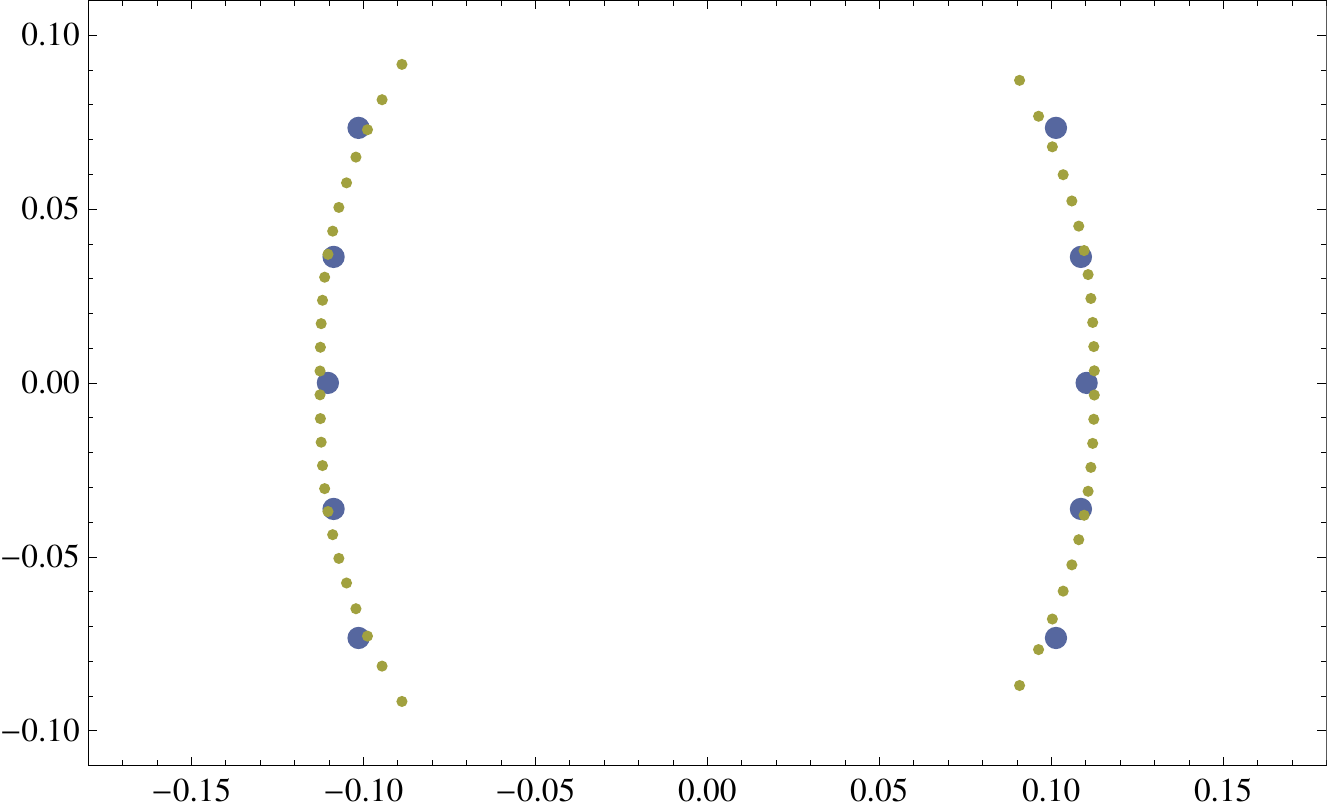}
\caption{Two Bethe roots configurations for a folded string with $\alpha=1/3$ and $L_1=30,150$, represented by the larger and smaller bullets respectively. The horizontal and vertical axes are the real and imaginary part of $u/L_1$, where $u$ is the rapidity of an excitation, related to its momentum by $u=\frac{1}{2} \cot \frac{p}{2}$. The case $L_1=30$ corresponds to the maximum number of roots that we used to compute $r_{1234}$ with formula \eqref{tail}. Even though this number of roots was not very large, we see that they lie nicely along the cuts formed by the clearly classical configuration $L_1=150$.
}
\label{foldedcuts}
\end{figure}
Having in mind the same setup considered in the previous paragraph, the goal of this subsection is to evaluate numerically our weak coupling formula \eqref{tail} by keeping $\alpha$ fixed and increasing the length of this operator. In order to evaluate \eqref{tail}, all we need are the Bethe roots of the two large operators.\footnote{For the light operators, recall that to obtain a BPS state in the spin chain language, we should send all rapidities to infinity, which is equivalent to have all excitations with zero momentum.} Since these are dual to the folded string, their Bethe roots are distributed along two symmetric cuts, as shown in figure \ref{foldedcuts}. In the setup we are considering, operator $\O_2$ has $J_1-4$ Bethe roots, while $\O_1$ has $J_1$ of them. It is then natural to ask where do we put the extra four roots of $\O_1$. Following the discussion in \cite{paper2}, we choose these four roots to lie on the already existing two symmetric cuts defined by the Bethe roots of $\O_2$. Given our setup, this means that we add two roots to each of the cuts. As we are about to see, this is the weak coupling choice that reproduces the strong coupling computation.

\begin{figure}[h]
\centering
\includegraphics[width=15cm]{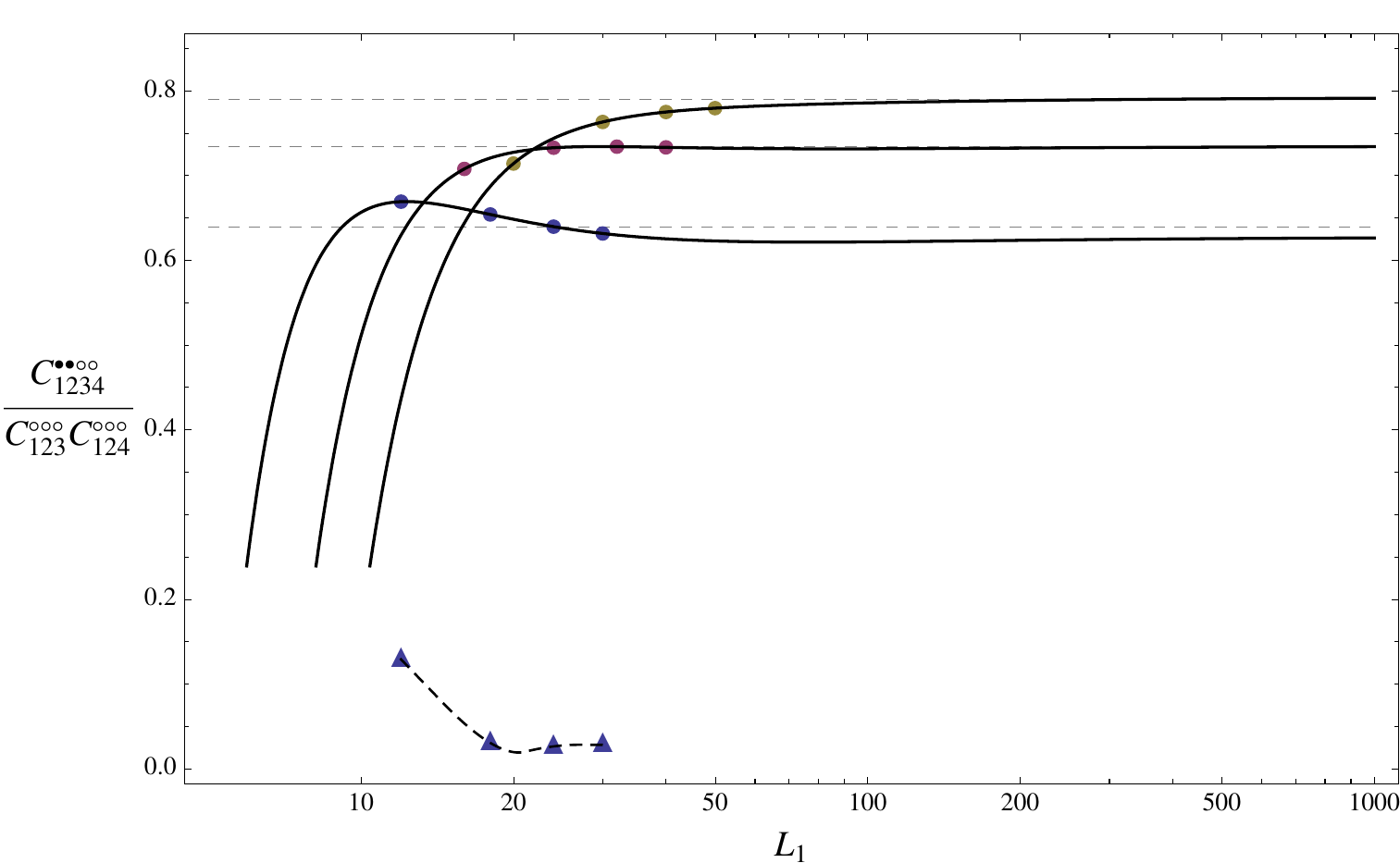}
\caption {The bullets correspond to the numerical data obtained by evaluating the weak coupling formula \eqref{tail} with the roots of $\O_1$ and $\O_2$ lying on the same classical cuts, for different values of $\alpha=1/3,1/4,1/5$, from bottom to top.  In this case the fits (solid black curves) perfectly match the analytical prediction from strong coupling (dashed gray lines) obtained from \eqref{intsigma}. The triangles correspond to the data obtained evaluating \eqref{tail} with $\alpha=1/3$ and by placing the extra roots of $\O_1$ at finite values outside the cuts of $\O_2$. Clearly, this case does not match the strong coupling prediction, confirming that the issue of back-reaction is also important for four-point functions.}
\label{dataplot}
\end{figure}

It is computationally easy to find the positions of a very large number of Bethe roots, see \cite{Bargheer:2008kj}. However, due to the sum over $l$, the many sums over partitions and the scalar products appearing in it, equation \eqref{tail} is quite non-trivial to evaluate numerically if we consider a large number of excitations.\footnote{Note that \eqref{tail} is quite more involved than the three-point function formula derived in \cite{paper1} and used in \cite{paper2}.} In practice we were only able to evaluate it for configurations with $4,6,8$ and $10$ Bethe roots in operator $\O_1$ using the Mathematica codes presented in appendix \ref{appcodes}. We collected this data in the following form $\{L_1,r_{1234}\}$, where $r_{1234}$ is the ratio on the l.h.s of \eqref{ratio1234}. For example, for filling fraction $\alpha=1/3$, we obtained
\begin{center}
{\small
\verb"data={{12,0.669138111344},{18,0.654008855014},{24,0.639613251186},"\\
\verb"     {30,0.631580997967}};"
}
\end{center}
We then fit this data in order to find the large $L_1$ asymptotics of $r_{1234}$. Since for a fixed $\alpha$ we only had four points of data, the fit we used was of the form $r_{1234} = a_0 + a_1/L_1+ a_2/L_1^2+a_3/L_1^3$.  Given the data for $\alpha=1/3$ shown above, we can simply run the following code in Mathematica
\begin{center}
{\small
\verb"Fit[data,1/L^Range[0,3],L]"
}
\end{center}
to obtain $a_0=0.627249$. In the second column of table \ref{tableratios} we present the value of $a_0$ for three different filling fractions. We see that, with very small error, the numerical weak coupling results approach the analytical strong coupling predictions in the Frolov-Tseytlin limit. Figure \ref{dataplot} shows our list of data points, their fits and the analytical predictions.
\begin{table}[h]
\beq\nn
\begin{array}{c|cccllcl}
\alpha & \text{Analytical prediction}& \text{Numerical extrapolation}&\text{Relative error} \\
\hline
 \frac{1}{3} & 0.639186 & 0.627249 & 1.868\% \\
 \frac{1}{4} & 0.734274 & 0.734578 & 0.042\% \\
 \frac{1}{5} & 0.789894 & 0.791947 & 0.259\%
\end{array}
\eeq
\caption{Numerical data obtained with the weak coupling formula \eqref{tail} (evaluated with the Bethe roots of $\O_1$ and $\O_2$ lying on the same classical cuts) compared with the analytical prediction from strong coupling \eqref{intsigma} for different values of the filling fraction of operator $\O_1$.}
\la{tableratios}
\end{table}

Recall from the discussion in \cite{paper2} that the other two choices for the positions of the extra roots of $\O_1$ are i) to put them at finite positions outside the existing cuts of $\O_2$ and ii) to put them at infinity. This is the issue of \textit{back-reaction} addressed in \cite{paper2}. The triangles in figure \ref{dataplot} represent the data points we obtained by evaluating \eqref{tail} with option i). As we can see, the extrapolation in this case does not match the strong coupling result. Hence, we confirmed numerically that the issue of back-reaction is also important for four-point functions.\footnote{Of course, we expect the case in which the extra roots are placed at infinity to also differ from the other two choices.}

Before closing this subsection, let us make some comments regarding the accuracy of our numerical results. By looking at table \ref{tableratios} and figure \ref{dataplot}, the skeptical reader might think that the results presented in this subsection are not conclusive enough to claim a weak/strong coupling match for four-point functions obeying the setup of figure \ref{setuppartitions}. However, we should note again that we only used four points of data to perform each fit, compared to the seven points used for the numerical weak/strong coupling match of three-point functions of two large and one small operators \cite{paper2}. In that case, the accuracy in the numerical match was indeed much higher. However, had we only considered the first four points of data obtained in \cite{paper2} to perform the fit for the three-point functions match, the relative error between the analytical and numerical results for $\alpha=1/3,1/4,1/5$ would have been $2.122\%, 0.612\%,0.381\%$, respectively. Hence, we see that the relative errors for the weak/strong coupling match of four-point functions shown in table \ref{tableratios} are in fact smaller than the relative errors in the match of three-point functions. We can be confident that adding more points of data to our fits would simply increase the accuracy of the numerical results even further, confirming that the weak/strong coupling match presented here is indeed exact.

\subsection{Analytics: four-point functions from coherent states}
Let us now prove analytically the match between weak and strong coupling for the four-point functions of interest.\footnote{We will follow closely the logic of section 3 of \cite{paper2}, to where we refer the reader for more details.} We will use an alternative approach to the one presented in section \ref{sectionweak} and used in the previous subsection. Namely, we will consider the operators $ \mathcal{O}_1 $ and $ \mathcal{O}_2 $ to be classical with their charges and their lengths taken to be much larger than the corresponding charges and lengths of $ \mathcal{O}_3 $ and $ \mathcal{O}_4 $. In this limit, the exact Bethe states (\ref{operators}) are well approximated by coherent states \cite{Fradkin}
\begin{eqnarray}
| \O_1 \>&=& \dots \otimes\left|{\bu}(\tfrac{n}{\mathcal{L}_1}) \right\> \otimes\left| {\bu}(\tfrac{n+1}{\mathcal{L}_1}) \right\>\otimes \dots \\
\< \O_2 |&=& \dots \otimes\left\<{\bbu}(\tfrac{n}{\mathcal{L}_2}) \right|\otimes \left\< {\bbu}(\tfrac{n+1}{\mathcal{L}_2}) \right| \otimes\dots
\end{eqnarray} 
where $\mathcal{L}_i \equiv L_i/2\pi$ and at each site $|\bold{u} \rangle = u_{1}|X\rangle +u_2 |Z\rangle$, such that $\<\bu|\bu\>=\bar\bu \cdot \bu=1$.\footnote{Note that to describe a given coherent state, we use the same notation $u_j$ used in the strong coupling discussion, see equations \eqref{ratio123} and \eqref{ratio124}. This is because these quantities will be matched between weak and strong coupling.}  We note that these states do not depend on the Bethe roots and are completely disentangled. The operators $ \mathcal{O}_3 $ and $ \mathcal{O}_4 $  are chosen to be vacuum descendant states, corresponding to the BPS operators
\begin{equation}
\mathcal{O}_3 = \mathcal{N}_3 \bigl(\Tr[\bar{X}^{j_1}Z^{j_2}] +\text{permutations} \bigr),
\end{equation} 
\begin{equation}
\mathcal{O}_4 = \mathcal{N}_4 \bigl(\Tr[\bar{X}^{k_1}Z^{k_2}]+\text{permutations} \bigr),
\end{equation}
where the normalization constants $ \mathcal{N}_3 $ and $\mathcal{N}_4 $ are one over the square root of the number of distinct permutations
\begin{equation}
\mathcal{N}_3=\sqrt{\frac{j_1!j_2!}{(j_1+j_2-1)!}} \, ,
\end{equation}
\begin{equation}
\mathcal{N}_4=\sqrt{\frac{k_1!k_2!}{(k_1+k_2-1)!}} \, .
\end{equation}

Now consider the setup of figure \ref{setuppartitions}. The operator $ \mathcal{O}_3 $ is Wick contracted with $ \mathcal{O}_1 $ and $ \mathcal{O}_2 $ at sites\footnote{Due to the periodicity of the spin chain, the sites $q=0,-1,-2,\dots$ correspond to $ L_1,L_1-1,L_1-2 \dots $} $ q,q-1,\dots $. For a given $ q $ and $ l $, the Wick contractions between the large operators $ \mathcal{O}_1 $ and $ \mathcal{O}_2 $ are given by
\begin{equation}\label{I}
\mathcal{I}_{q,l}=\prod_{i=q+1}^{l+q} \bbu(i/\mathcal{L}_2)\cdot\bu(i/\mathcal{L}_1) \prod_{j=q+l+k_1+1}^{l_{12}+q+k_1} \bbu(j/\mathcal{L}_2)\cdot\bu(j/\mathcal{L}_1) \, .
\end{equation}
Let us consider the contractions between $ \mathcal{O}_3 $ or $ \mathcal{O}_4 $ with the classical operators. Given the present configuration, the only term of $ \mathcal{O}_3 $ which gives a nonzero contribution is $ \Tr[Z^{j_2}\bar{X}^{j_1}] $ and similarly for $ \mathcal{O}_4 $. For a fixed $ q $ and $ l $, we denote the Wick contractions of $ \mathcal{O}_3 $ and $\mathcal{O}_4 $ with the classical operators by $ \mathcal{J}^{(3)}_{q,l} $ and $ \mathcal{J}^{(4)}_{q,l} $, respectively. We have approximately
\beqq\label{J3}
\mathcal{J}^{(3)}_{q,l}\simeq \mathcal{N}_3\,u_{1}^{j_1}\left(\frac{q}{\mathcal{L}_1} \right)\,\bar{u}_{2}^{j_2}\left(\frac{q}{\mathcal{L}_2} \right) \, ,
\eeqq
\beqq\label{J4}
\mathcal{J}^{(4)}_{q,l}\simeq \mathcal{N}_4\,u_{1}^{k_1}\left(\frac{q+l+1}{\mathcal{L}_1} \right)\,\bar{u}_{2}^{k_2}\left(\frac{q+l+1}{\mathcal{L}_2} \right) \, .
\eeqq
Finally, we get\footnote{The upper limit in the sum over $l$ should be $L_1-j_1-k_1$. We replaced it by $ L_1 $, since $ L_1 \gg j_1,k_2 $.}
\begin{equation}\label{c1234}
C_{1234}^{\bullet\bullet\circ\circ}\simeq \sum_{l=0}^{L_1}\sum_{q=0}^{L_1} \mathcal{I}_{q,l} \, \mathcal{J}^{(3)}_{q,l}  \mathcal{J}^{(4)}_{q,l} \, .
\end{equation}

Following \cite{paper2}, we will now argue that $ \mathcal{I}_{q,l} $ is one to leading order. For that we may use a gauge invariance of the absolute value of $ C_{1234}^{\bullet\bullet\circ\circ} $: by multiplying each single site state by a phase $ |\bu^{(j)} \> \rightarrow e^{-i\phi(j)}|\bu^{(j)}\> $, $ |C_{1234}^{\bullet\bullet\circ\circ}| $ must be invariant. We fix this gauge by choosing a conformal-like gauge, resembling the Virasoro constraint in the string theory side, which reads\footnote{In the continuum limit, we introduce the variable $\sigma=2\pi \frac{n}{L}$ and $\bu^{(n)}\rightarrow \bu(\sigma)$.} $\bbu\cdot \partial_{\sigma}\bu = 0$. Then, with this gauge and using 
\begin{equation}\label{aprox}
\mathcal{L}_2=\mathcal{L}_1+\frac{k_1+j_1-k_2-j_2}{2\pi} \, ,
\end{equation} 
\eqref{I} can be rewritten as
\begin{eqnarray*}
 \mathcal{I}_{q,l} &  \simeq & \exp \int\limits_{0}^{2\pi l} \, \frac{d\sigma}{2\pi}\, \log \(\bbu \cdot \[ \bu - \frac{j_2+k_2-j_1-k_1}{2\pi \mathcal{L}_2} \sigma \partial_{\sigma} \bu \] \)\,\\
& & \times \exp \int\limits_{0}^{2\pi (l_{12}-l)} \,\frac{d\sigma'}{2\pi}\, \log \(\bbu \cdot \[ \bu - \frac{j_2+k_2-j_1-k_1}{2\pi \mathcal{L}_2} \sigma' \partial_{\sigma'}\bu \] \) \\
&\simeq & 1 + \mathcal{O}\(\frac{1}{\mathcal{L}_2} \).
\end{eqnarray*} 
With this simplification, the remaining part of the formula becomes
\begin{equation}\label{finalform}
\begin{split}
C_{1234}^ {\bullet\bullet\circ\circ}&\simeq  \,\, \mathcal{N}_3 \, \mathcal{N}_4  \, \sum_{l=0}^{L_1}\sum_{q=0}^{L_1}  u_{1}^{j_1}\left(\frac{q}{\mathcal{L}_1} \right)\,\bar{u}_{2}^{j_2}\left(\frac{q}{\mathcal{L}_2} \right) u_{1}^{k_1}\left(\frac{q+l+1}{\mathcal{L}_1} \right)\,\bar{u}_{2}^{k_2}\left(\frac{q+l+1}{\mathcal{L}_2} \right) \\
& \simeq \sqrt{\frac{j_1!j_2!}{(j_1+j_2-1)!}} \, \sqrt{\frac{k_1!k_2!}{(k_1+k_2-1)!}} \, \int\limits_0^{2\pi} L_1 \frac{d\sigma}{2\pi} u_{1}^{j_1}(\sigma) \bar{u}_{2}^{j_2}(\sigma) \int\limits_0^{2\pi} L_1 \frac{d\sigma'}{2\pi} u_{1}^{k_1} (\sigma')\bar{u}_{2}^{k_2}(\sigma')
\end{split}
\end{equation}
where in the second line we have used (\ref{aprox}) and the periodicity of the chain, which allowed us to factorize the double sum in the first line. Upon dividing this expression by $C_{123}^{\circ\circ\circ} C_{124}^{\circ\circ\circ}$, which can be read from \eqref{bpssimp}, we exactly obtain the strong coupling result \eqref{ratio1234}!


\section{More general four-point functions} 
\begin{figure}
\centering
\def\svgwidth{10.15cm}
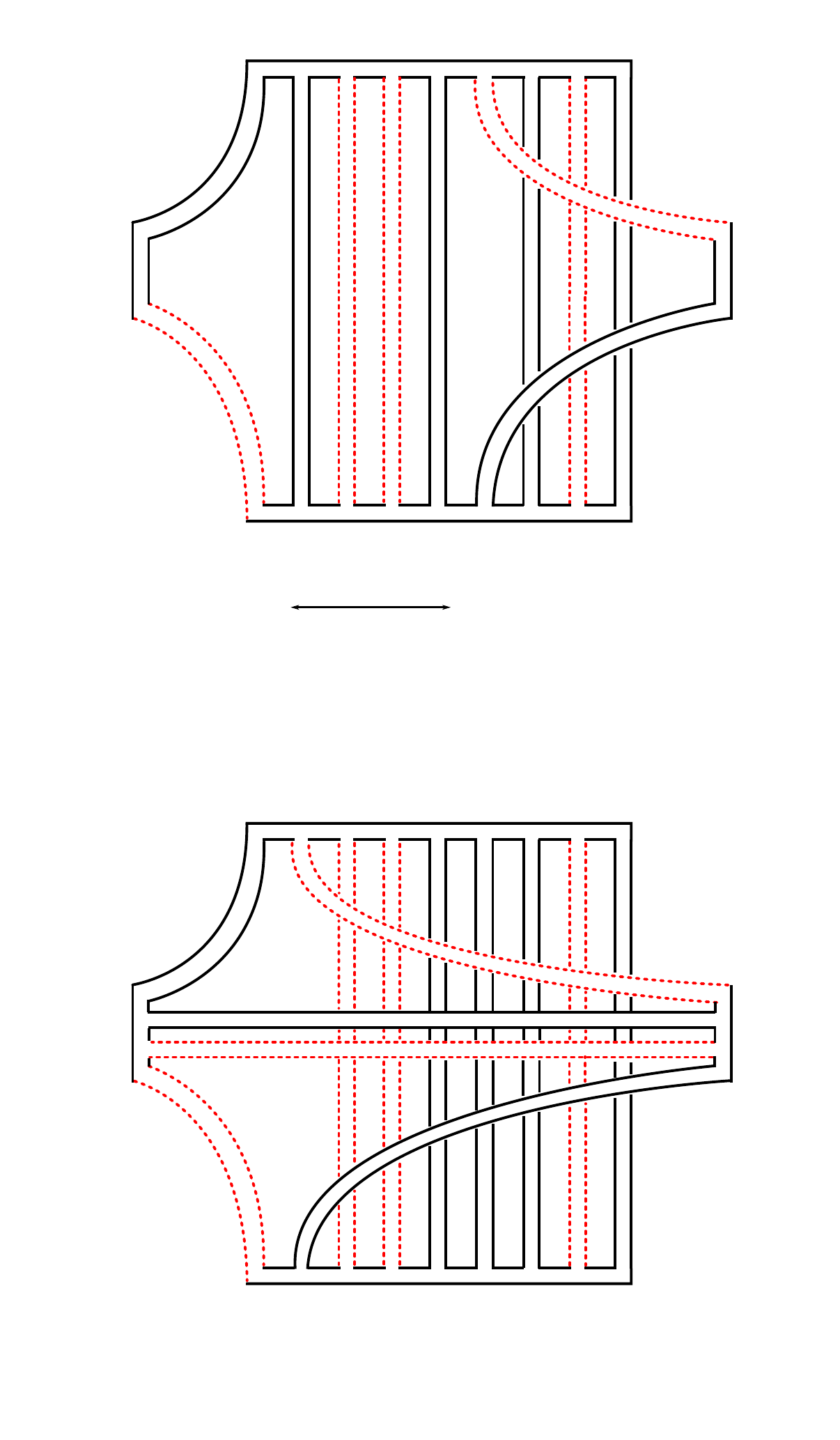
\caption{More general four-point functions in the $SU(2)$ sector. (a) It is clear that the diagrams with no contractions between $\O_3$ and $\O_4$ are planar for any $l \geq 0$. (b) If there are contractions between $\O_3$ and $\O_4$, the only planar diagrams are those in which $l=0$.  When $\O_3$ and $\O_4$ are taken to be much smaller than $\O_1$ and $\O_2$, this weak coupling configuration is analogous to the holographic four-point function depicted in figure \ref{2light}.}
\label{lcontract}
\end{figure}

\la{anothersetup}
In the case we studied in the previous sections, operators $\O_3$ and $\O_4$ did not contract between themselves, see figure \ref{setuppartitions}. Such a configuration was motivated by the fact that at strong coupling these were taken to be light operators and their interactions are suppressed. Indeed, in the tree-level diagram of figure  \ref{2light} only a single integration over the worldsheet is performed and hence it is suppressed by $1/L_1$ with respect to the diagram of figure \ref{4ptstrongcoupling}.\footnote{By $1/L_1$ supressed we simply mean that the case of figure \ref{4ptstrongcoupling} is of order $\O(L_1^2)$, while that of figure \ref{2light} goes as $\O(L_1)$ precisely due to the number of integrations over the worldsheet involved in each case.}. We can therefore address the question of whether or not we still have a match with the strong coupling result when we consider a weak coupling setup such as the one shown in figure \ref{lcontract}, in which the light operators can interact. In particular, is the strong coupling suppression mentioned above manifest at weak coupling? 
\begin{figure}[t]
\centering
\def\svgwidth{11.5cm}
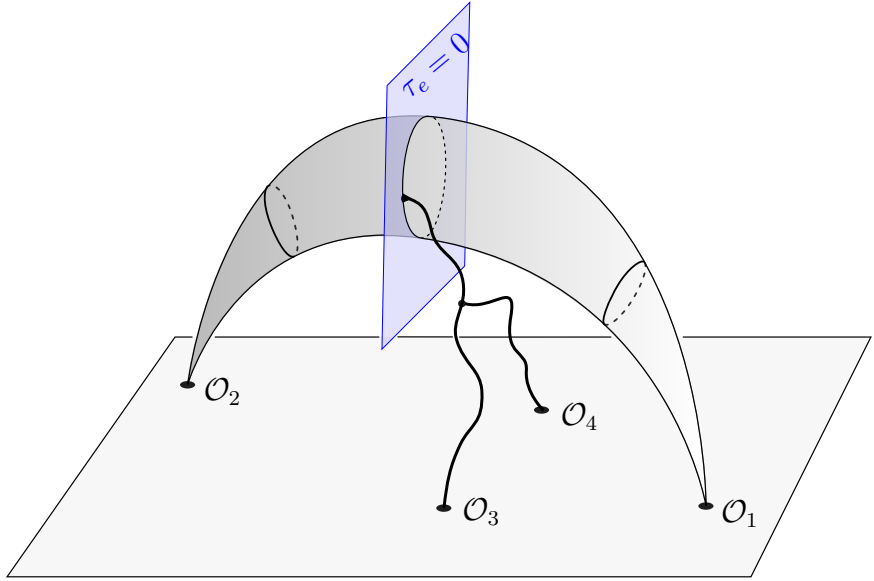
\caption{Although this is a tree-level diagram, it is suppressed by $1/L_1$ with respect to the case of figure \ref{4ptstrongcoupling} since in this case only one integration over the worldsheet is performed. This is analogous to the weak coupling configuration of figure \ref{lcontract}b.}
\la{2light}
\end{figure}

For the setup we consider in this section we need to specify the number of excitations ($X$ and $\bar X$) and vacuum fields ($Z$ and $\bar Z$) contracted between operators $\O_3$ and $\O_4$. We will denote these by $m_1$ and $m_2$, respectively. In this case the four-point function reads\footnote{We recall that we are excluding the disconnected diagrams that may appear when $\{j_1,j_2\}=\{k_1,k_2\}$.}
\begin{equation}
\begin{split}
G_{4}(x_1,x_2,x_3,x_4)=\frac{1}{N^2}\sum_{m_1=0}^{\min{\{j_1,k_1\}}}\sum_{m_2=0}^{\min{\{j_2,k_2\}}}C_{1234,\{m_1,m_2\}}\,\mathcal{G}_{\{m_1,m_2\}}(x_1,x_2,x_3,x_4)
\end{split}
\end{equation}
where\begin{equation}\begin{split}
\mathcal{G}_{\{m_1,m_2\}}(x_1,x_2,x_3,x_4)=&\,\frac{1}{|x_{12}|^{2(L_1-j_2-k_1+m_1+m_2)} |x_{13}|^{2(j_2-m_2)} |x_{14}|^{2(k_1-m_1)} } \, \times \\
&\times \frac{1}{|x_{23}|^{2(j_1-m_1)} |x_{24}|^{2(k_2-m_2)} |x_{34}|^{2(m_1+m_2)}} \,.
\end{split}\end{equation}
We will now compute $G_{4}(x_1,x_2,x_3,x_4)$ for the case where $\O_1$ and $\O_2$ are heavy operators and $\O_3$ and $\O_4$ are light chiral primary operators, using the coherent state language. In appendix \ref{genint}, equation \eqref{tail2}, we provide the formula for $G_{4}(x_1,x_2,x_3,x_4)$ obtained using the integrability tools, valid for any four operators obeying the setup of figure \ref{lcontract}.

In the classical limit, we again impose the conformal-like gauge such that the contractions between heavy operators are approximately one. The contributions from the contractions between the light operators $\mathcal{O}_3$ and $\mathcal{O}_4$ with the heavy operators are respectively given by 
\beqq
\mathcal{J}^{(3)}_{q,l,m_1,m_2} \simeq \binom{m_1+m_2}{m_1}\,\mathcal{N}_3\,\bar{u}_{1}^{j_1-m_1}\left(\frac{q}{\mathcal{L}_2} \right)\,u_{2}^{j_2-m_2}\left(\frac{q}{\mathcal{L}_1} \right) \, ,
\eeqq
\beqq
\mathcal{J}^{(4)}_{q,l,m_1,m_2}\simeq \binom{m_1+m_2}{m_1}\,\mathcal{N}_4\,u_{1}^{k_1-m_1}\left(\frac{q+l+1}{\mathcal{L}_1} \right)\,\bar{u}_{2}^{k_2-m_2}\left(\frac{q+l+1}{\mathcal{L}_2} \right) \, ,
\eeqq
where $q$ is the site of the heavy operators where $\mathcal{O}_3$ is inserted and the binomial factor $\binom{m_1+m_2}{m_1}$ is the number of contractions between the light and heavy operators.

For $ l\neq 0 $ and $ m_1+m_2\neq0 $ the diagrams are non-planar, and hence $ 1 / N $ suppressed. Hence, in the planar limit, the four-point function is given by
\begin{equation}
\begin{split}
G_4(x_1,x_2,x_3,x_4)\,&\simeq  \, \frac{1}{N^2}\sum_{l=0}^{L_1}\sum_{q=0}^{L_1}\,\mathcal{G}_{\{0,0\}}(x_1,x_2,x_3,x_4)\,\mathcal{J}^{(3)}_{q,l,0,0} \,\mathcal{J}^{(4)}_{q,l,0,0}\,+\\ &+\frac{1}{N^2}\sum_{m_1=0}^{\min{\{j_1,k_1\}}}\sum_{m_2=0}^{\min{\{j_2,k_2\}}}\sum_{q=0}^{L_1}\, \delta_{m_1+m_2\neq 0}\,\mathcal{G}_{\{m_1,m_2\}}(x_1,x_2,x_3,x_4)\,\mathcal{J}^{(3)}_{q,0,m_1,m_2}\mathcal{J}^{(4)}_{q,0,m_1,m_2}
\end{split}
\end{equation}
where the first term takes into account the contribution of the diagrams for which $\{m_1,m_2\}=\{0,0\}$ and any $l$. The second term corresponds to the diagrams with $l=0$ and $m_1+m_2\neq 0$. Taking the continuum limit of this expression, we obtain
\begin{equation}
\begin{split}\label{light2}
G_4(x_1,x_2,x_3,x_4)\, \simeq \, & \,\frac{1}{N^2}\,\mathcal{G}_{\{0,0\}}(x_1,x_2,x_3,x_4)\, \mathcal{N}_3 \, \mathcal{N}_4 \int\limits_0^{2\pi} L_1 \frac{d\sigma}{2\pi} \bar{u}_{1}^{j_1}(\sigma) u_{2}^{j_2}(\sigma) \int\limits_0^{2\pi} L_1 \frac{d\sigma'}{2\pi} u_{1}^{k_1} (\sigma')\bar{u}_{2}^{k_2}(\sigma')\, + \\ 
&+\frac{1}{N^2}\,\sum_{m_1=0}^{\min{\{j_1,k_1\}}}\sum_{m_2=0}^{\min{\{j_2,k_2\}}} \delta_{m_1+m_2\neq 0}\,\mathcal{G}_{\{m_1,m_2\}}(x_1,x_2,x_3,x_4)\,\binom{m_1+m_2}{m_1}^{2}\,\mathcal{N}_3 \, \mathcal{N}_4\, \times \\
&\times \int\limits_0^{2\pi} L_1 \frac{d\sigma}{2\pi} \bar{u}_{1}^{j_1-m_1}(\sigma) u_{2}^{j_2-m_2}(\sigma)  u_{1}^{k_1-m_1} (\sigma)\bar{u}_{2}^{k_2-m_2}(\sigma) \, ,
\end{split}
\end{equation} 
The second term is suppressed by $ 1/L_1 $ compared to the first one and thus, we recover the strong coupling result \eqref{ratio1234}. In this way, we made manifest that the $1/L_1$ suppression due to the interaction of light modes in the string theory side, see figure \ref{2light}, has a dual description in the gauge theory side in the classical limit. It would be very interesting to compare the second term in \eqref{light2} with the result of a strong coupling computation of the tree-level four-point function of figure \ref{2light} and see if a match occurs for these results.

Finally, note that in order to evaluate \eqref{light2}, we need to specify the charges of the four operators according to \eqref{Vacchoicestrong}, the values of $m_1,m_2$ and the variables $u_1,u_2$. The latter depend on the classical string solution dual to the heavy operators $\O_1$ and $\O_2$ (e.g.\ see section 3 of \cite{paper2} for their explicity expressions when the heavy operators are dual to the folded string).


\section{Conclusions and discussion} \la{conclusions}
This paper was devoted to the study of four-point functions of single trace gauge-invariant operators in the $SU(2)$ sector of $\mathcal{N}=4$ SYM. By representing each operator as an $SU(2)$ spin chain state, we applied the integrability techniques introduced in \cite{paper1} which allowed us to solve the problem to leading order at weak coupling for \textit{any} four operators obeying the setup of figure \ref{setuppartitions}. 

We then studied the classical limit of such correlation functions, focusing on the case in which the charges of two of the operators are taken to be much larger than those of the other two. We provided numerical evidence for the match of this weak coupling result with the strong coupling computation in the Frolov-Tseytlin limit. Furthermore, we managed to explain analytically this match, using spin chain coherent states to describe the two heavy operators. 

An important aspect for the numerical match was related to the difference between operators $\mathcal{O}_1$ and $\mathcal{O}_2$. Given that operators $\mathcal{O}_3$ and $\mathcal{O}_4$ were taken to be small, it is reasonable to expect that $\mathcal{O}_1$ should be approximately the complex conjugate of $\mathcal{O}_2$. In the classical limit, the Bethe roots of operator $\mathcal{O}_2$ organize into cuts. The slight difference between $\mathcal{O}_1$ and the complex conjugate of $\mathcal{O}_2$ comes from the position of the few extra roots of $\O_1$ with respect to those of $\O_2$. Given the cuts defined by the roots of $\O_2$, there are three possibilities for where to put the extra roots of $\O_1$: they can be placed on the existing cuts, at finite position outside the existing cuts, or at infinity. Based on a similar analysis for the three-point function problem \cite{paper2}, we chose the first option to obtain the referred weak/strong coupling match. We also showed that, just as in the case of three-point functions, the issue of back-reaction is important for four-point functions. That is, the weak coupling results obtained using equation \eqref{tail} are sensitive to the different choices for the positions of the extra roots of $\O_1$ in such a way that only the choice we made for them matches with the strong coupling result \eqref{ratio1234}. It would be very interesting to study the issue of back-reaction at strong coupling.

We also considered a more general class of four-point functions in the $SU(2)$ sector of the theory by allowing contractions among all four operators. When $\mathcal{O}_3$ and $\mathcal{O}_4$ are light operators, we found two types of contributions at weak coupling: the diagrams in figure \ref{lcontract}a give the leading contribution, while those in figure \ref{lcontract}b are suppressed in the length of the heavy operators. Furthermore, by representing the two heavy operators by coherent states, we computed the exact form of these supressed terms. At strong coupling, there is an analogous behaviour. Two possible contributions at leading order are represented in figures \ref{4ptstrongcoupling} and \ref{2light}, with the former being the dominant  one and the latter being suppressed in the length of the heavy operators since a single integration is performed over the worldsheet. It would be very interesting to perform the strong coupling computation of the tree-level holographic four-point function depicted in figure \ref{2light} and see if a match between weak and strong coupling occurs in this case.

Also, using integrability techniques, we showed how to compute the four-point function of figure \ref{lcontract} to leading order at weak coupling for operators of \textit{arbitrary size}, see \eqref{tail2}. We expect this formula and also \eqref{light2} to capture part of the result for the one-loop four-point function at strong coupling.\footnote{As pointed out in \cite{paper1}, we need to do two things in order to get one-loop data from the tree-level formulas obtained using integrability. Firstly, we need to consider the $\O(\lambda)$ correction to the Bethe wave function of the operators, obtained by diagonalizing the two-loop Hamiltonian. Secondly, we have to consider Hamiltonian insertions \cite{Okuyama:2004bd,Roiban:2004va,Alday:2005nd} when performing the contractions between all operators.} However, it would be important to first check this claim in the case of three-point functions, see the discussion in \cite{paper1, paper2}.

We should stress that our weak coupling formulas \eqref{tail} and \eqref{tail2} are valid for any four operators in the $SU(2)$ sector . In particular, once an equivalent computation is performed at strong coupling\footnote{Recently there has been progress towards tackling the computation of general holographic three-point functions at strong coupling \cite{paper2,Vicedo,KloseMcL}.}, it would be interesting to make a comparison of both results for the cases when more than two operators in the four-point function are heavy and see if a weak/strong coupling match occurs.
 
The strong coupling results presented in section \ref{sectionstrong} can be generalized to the $SU(3)$ sector \cite{paper2}. The same is not true at weak coupling, where the necessary integrability tools are not yet known, but such problem is currently under investigation \cite{toappear2}.
Nevertheless, the match of the four-point functions of two large and two small operators persists in the $SU(3)$ sector. This can be easily seen using $SU(3)$ coherent states as in \cite{paper2}.

Finally, it was recently reported \cite{Georgiou:2011qk} that the match of tree-level structure constants in the classical limit is also valid in the $SL(2)$ sector of the theory.  It would be interesting to use the integrability tools for the $SL(2)$ sector presented in appendix A of \cite{paper1} to see whether or not the match for four-point functions and the issue of back-reaction also extend to that sector.

\section*{Acknowledgments}
We thank P. Vieira for suggesting the problem and for useful discussions. We thank M. Costa, N. Gromov, J. Penedones and A. Sever for comments. We thank J. Rodrigues and CFP for computational facilities. JC is funded by the Funda\c{c}\~ao para a Ci\^encia e Tecnologia fellowship SFRH/BD/69084/2010. This work has been supported in part by the Province of Ontario through ERA grant ER 06-02-293. Research at the Perimeter Institute is supported in part by the Government of Canada through NSERC and by the Province of Ontario through MRI. This work was partially funded by the research grants PTDC/FIS/099293/2008 and CERN/FP/116358/2010.
\emph{Centro de F\'{i}sica do Porto} is partially funded by FCT under grant PEst-OE/FIS/UI0044/2011. 


\appendix

\section{Cutting, flipping and sewing spin chains}
\la{tailoringtools}

In this appendix, we quickly summarize the integrability tools introduced in \cite{paper1} for the computation of correlation functions of single trace gauge-invariant operators in the $SU(2)$ sector of $\N=4$ SYM.  Note that all formulas we present in this appendix are in the coordinate Bethe ansatz base (see \cite{paper1} for details).

\subsection*{Cutting}
Consider a closed $SU(2)$ spin chain described by a Bethe state parametrized by a set of $N$ Bethe roots $\{u\}$. We can break it into a left and a right open subchain of lengths $l$ and $r$ as
\beq\la{Cutting}
|\{u\}\>=\sum_{\alpha\cup\bar\alpha =\{u\}} e_l^{\bar\alpha} \, {f^{\alpha\bar\alpha}f_<^{\bar\alpha\bar\alpha}f_<^{\alpha\alpha}\over f_<^{\{u\}\{u\}}} \, |\alpha\>_l\otimes|\bar\alpha\>_r \, ,
\eeq
where the sum runs over all $2^N$ possible ways of splitting the rapidities into two partitions $\alpha$ and $\bar\alpha$. For example, if $N=2$, the possible partitions $\( \a,\bar \a \)$ would be $\( \{\},\{u_1,u_2\} \),$ $\( \{u_1\},\{u_2\} \),  \( \{u_2\},\{u_1\} \),  \( \{u_1,u_2\}, \{\} \)$. We are also using the following shorthand notation
\beq
e_l^{\bar\alpha} = \prod_{u_j \in \bar\alpha} \({u_j+{i\over2}\over u_j-{i\over2}}\)^l \, ,\qquad  f^{\alpha\bar\alpha} \equiv \!\!\! \prod_{\scriptsize \begin{array}{c} {u_i} \in \alpha\\ v_j \in \bar \alpha\end{array}} \!\!\!f(u_i-v_j)\ ,\qquad f_<^{\alpha\alpha} \equiv \!\!\! \prod_{\scriptsize \begin{array}{c} {u_i, u_j} \in \alpha  \\  i<j \end{array}} \!\!\!f(u_i-u_j) \, 
\la{usefulnotation}
\eeq
and
\beq
f(u) \equiv 1+ \frac{i}{u} \, .
\eeq

\subsection*{Flipping}
Consider again a Bethe state on a spin chain of length $L$. We can flip the state as follows:
\beq
{\cal F}\circ |\{u\}\> =e^{\{u\}}_{L+1} \, {f_>^{\{u\}\{u\}}\over f_<^{\{u\}\{u\}}} \,
\<\{u^*\}|\hat C \, ,
\la{Flipping}
\eeq
where $\hat C$ stands for a charge conjugation which exchanges $Z\leftrightarrow\bar Z$ and $X \leftrightarrow\bar X$ in the operator language. We are using a shorthand notation in the same spirit as equations \eqref{usefulnotation}, where we would simply need to replace $\alpha$ by $\{u\}$.

\subsection*{Sewing}
Consider two Bethe states on spin chains of length $L$ parametrized by their Bethe roots $u_i$ and $v_i$, with $i=1,\dots,N$. The scalar product $S_N (\{v\},\{u\}) \equiv \< \{v^*\} | \{u\} \>$ can be efficiently computed using the new recursion relation for $SU(2)$ scalar products derived in \cite{paper1}. It reads
\beqa
&&S_N \(\{v_1,\dots,v_N\},\{u_1,\dots,u_N\}\) = \Biggl[ \sum_{n} b_n \, S_{N-1} \(\{v_1,\dots, \hat v_n,\dots,v_N\},\{\hat u_1, u_2,\dots,u_N\}\) \Bigr. \nn  \\
&&- \Biggl. \sum_ {n<m}  c_{n,m} \, S_{N-1}\(\{ u_1,v_1,\dots \hat v_{n},  \dots , \hat v_{m},\dots v_N \}, \{\hat u_1,u_2,\dots,u_N\}\) \Biggr] \text{AtoC}\(\{u\},\{v\}\)   ,
\la{newrecco}
\eeqa
where a hatted Bethe root means that it is omitted and
\beqq
\text{AtoC}\(\{u\},\{v\}\) = \frac{1}{d^{\{u\}} a^{\{v\}} g^{\{u+\frac{i}{2}\}} g^{\{v-\frac{i}{2}\}} f_<^{\{u\}\{u\}} \, f_>^{\{v\}\{v\}}} \, ,
\eeqq
\beqq
b_n = g(u_1-v_n) a(v_n) d(u_1) \prod_{j\neq n}^N f(u_1-v_j) f(v_j-v_n)  + \(u_1 \leftrightarrow v_n\) , 
\eeqq
\beqq
c_{n,m} = {g(u_1-v_{n}) \, g(u_1-v_{m}) \, a(v_{m}) d(v_{n})}{f(v_{n}-v_{m})} \prod_{j\neq n,m}^N  f(v_{n} - v_j)  \, f(v_j - v_{m}) + \(n \leftrightarrow m\) , 
\eeqq
where
\beq
g(u) \equiv \frac{i}{u} \, , \qquad a(u) \equiv \(u+\frac{i}{2}\)^L \, , \qquad d(u) \equiv \(u-\frac{i}{2}\)^L \, .
\eeq

Equation \eqref{newrecco} provides a complete solution for any $SU(2)$ scalar product and is the one we used to perform the explicit numerical evaluation of the four-point function at weak coupling, see equation \eqref{tail}. However, when one of the states in the scalar product is a BPS state, the formula for the scalar product simplifies dramatically \cite{paper1}. Recall that in the spin chain language a BPS state is that in which all excitations have zero momentum or, equivalently, infinity rapidity. Hence, if we send the Bethe roots $\{v\}$ to infinity in the scalar product above, we obtain the following simpler formula:
\beq\la{scalarBPS}
\<\{\infty\}|\{u\}\>={\,(-1)^N \, N!\over  g^{\{u+\frac{i}{2}\}}f_<^{\{u\}\{u\}}}\sum_{\alpha\cup\bar\alpha=\{u\}}(-1)^{|\alpha|} \,
e^{\alpha}_L \, f^{\bar\alpha\alpha}~.
\eeq


\section{Four-point functions of section \ref{anothersetup} using integrability}
\la{genint}

We compute $G_4(x_1,x_2,x_3,x_4)$ for the setup of figure \ref{lcontract} using the integrability tools of appendix \ref{tailoringtools}. The steps we need to follow are very similar to the ones outlined for the setup of figure \ref{setuppartitions}, see section \ref{sectionweak}, except that now we also have to cut operators $\O_3$ and $\O_4$ more than once and perform some extra scalar products.
\begin{figure}[h]
\centering
\def\svgwidth{11.0cm}
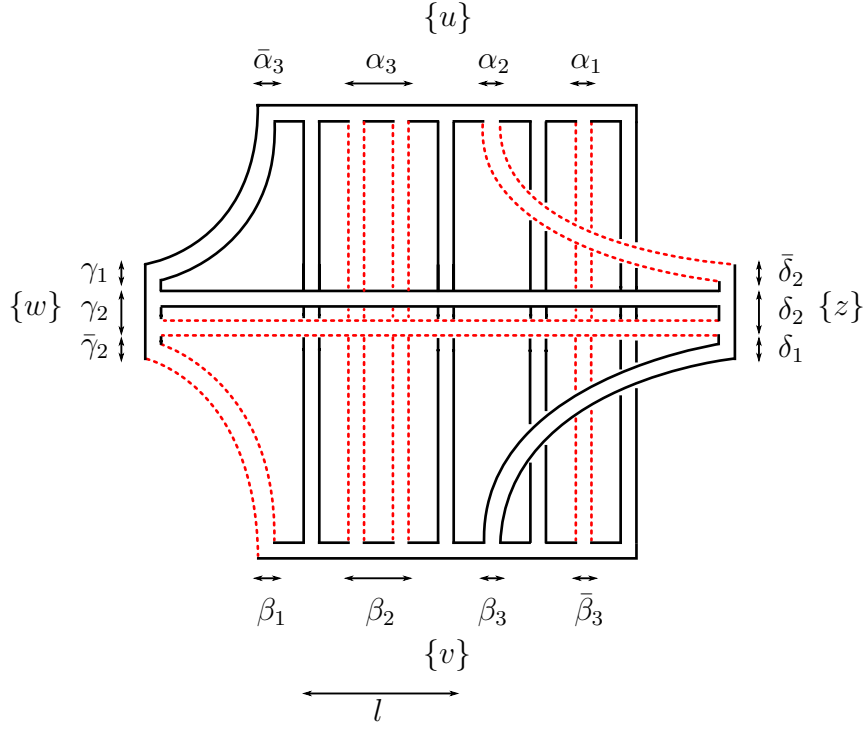
\caption{Setup for the computation of the four-point functions of section \ref{anothersetup} at weak coupling. The black (solid) lines represent vacuum fields, while the red (dashed) lines represent excitations. The figure shows the labelling of the excitations of each operator and the different partitions needed to perform the Wick contractions.}
\la{lightint}
\end{figure}
In the planar limit, $G_4(x_1,x_2,x_3,x_4)$ is given by
\begin{equation}\begin{split}\label{full}
G_4(x_1,x_2,x_3,x_4)\,=&\,\frac{1}{N^2}\,\mathcal{G}_{\{0,0\}}(x_1,x_2,x_3,x_4)\sum_{l=0}^{L_1 - L_3 + j_1 - k_1}C_{1234,\{0,0,l\}}\,\,+ \\
&+\,\frac{1}{N^2}\sum_{m_1=0}^{\min{\{j_1,k_1\}}}\sum_{m_2=0}^{\min{\{j_2,k_2\}}}\delta_{m_1+m_2\neq 0}\,\mathcal{G}_{\{m_1,m_2\}}(x_1,x_2,x_3,x_4)\,C_{1234,\{m_1,m_2,0\}} \, ,
\end{split}\end{equation}
where
\begin{align}
C_{1234,\{m_1,m_2,l\}}=& \,\, \Omega \,
\sum_{\a_1 \cup \bar{\a}_1 =\{u\}} \sum_{\a_2 \cup \bar{\a}_2 = \bar{\a}_1} \sum_{\beta_1 \cup \bar{\beta}_1 = \{v\}} \sum_{\substack{\beta_2 \cup \bar{\beta}_2 = \bar\beta_1 \\ |\beta_2| = |\a_3| \\ |\bar{\beta}_2| = |\a_1 |}} \sum_{\substack{\gamma_2 \cup \bar{\gamma}_2 = \{w\} \\ |\bar\gamma_2| = |\beta_1|}} \sum_{\substack{\delta_2 \cup \bar{\delta}_2 = \{z\} \\ |\bar\delta_2| = |\a_2|}} \nn \\
&\times \, e^{\bar\a_1}_{L_1 -L_3+j_1 -k_1+m_1+m_2-l} \, e^{\bar\a_2}_{k_1-m_1} \, \frac{f^{\a_1 \bar\a_1} f^{\a_1 \a_1}_{<} f^{\a_2 \bar\a_2} f^{\a_2 \a_2}_{<} f^{\bar\a_2 \bar\a_2}_<}{f^{\{u\} \{u\}}_{<}} \nn \\
&\times \, e^{\bar\beta_1}_{j_1-m_1} \, e^{\beta_2}_{l+1} \, e^{\bar{\beta}_2}_{L_2-j_1+m_1+1} \, \frac{f^{\beta_1 \bar{\beta}_1} f^{\beta_1 \beta_1}_{<} f^{\beta_2 \bar\beta_2} f^{\beta_2\beta_2}_> f^{\bar{\beta}_2 \bar{\beta}_2}_{>}}{f^{\{v\} \{v\}}_{<}}  \nn \\
&\times \, e^{\{w\}}_{L_3-j_1-m_2} \, e^{\bar\gamma_2}_{j_1+m_2+1} \, \frac{f^{\gamma_2 \bar\gamma_2} f^{\gamma_2 \gamma_2}_< f^{\bar\gamma_2 \bar\gamma_2}_>}{f^{\{w\}\{w\}}_<} \, e^{\{z\}}_{L_4-k_1-m_2} \, e^{\delta_2}_{m_1+m_2+1} \, e^{\bar\delta_2}_{k_1+m_2+1} \, \frac{f^{\delta_2 \bar\delta_2} f^{\delta_2 \delta_2}_> f^{\bar\delta_2 \bar\delta_2}_>}{f^{\{z\}\{z\}}_<} \nn \\
&\times \, \< \beta_2^{\ast} | \a_3 \> \< \bar{\beta}_3^{\ast}| \a_1 \> \< \bar\gamma_2^* | \beta_1 \> \<\bar{\delta}_2^{\ast} | \a_2 \> \<\delta_2^* | \gamma_2 \> \, ,
\la{tail2}
\end{align}
where $\Omega$ is given in \eqref{omega} and the labeling for the Bethe roots is indicated in figure \ref{lightint}.\footnote{Just like for the setup of section \ref{sectionweak}, we can also include a symmetry factor in $\Omega$ to take into account the case when $\O_3$ or $\O_4$ are dropped from figure \ref{lcontract}. In this case, it reads
\beqq
\Omega_{m_1,m_2}=\sqrt{\frac{L_1 L_2 L_3^{\delta_{L_3>0}} L_4^{\delta_{L_4>0}}}{\N_1 \N_2 \N_3 \N_4}} \, \frac{1}{(L_1 - L_3 + j_1 - k_1+m_1+m_2)\bigl(\Theta(L_3)\Theta(L_4)-\delta_{L_3>0}\delta_{L_4>0}\bigr)+1} \, .
\la{omegam1m2}
\eeqq}
The first term in \eqref{full} corresponds to the contribution of diagrams for which $m_1,m_2=0$ and $l \geq 0$. In this case, expression (\ref{tail2}) simplifies since the partitions $\delta_2$ and $\gamma_2$ become empty. The second term in (\ref{full}) takes into account the contribution of planar diagrams where the light operators can contract and therefore we should have $l=0$. Equation \eqref{tail2} is valid for \textit{any} four operators obeying the setup presented in figure \ref{lcontract}. 


\section{Mathematica codes} \la{appcodes}
In this appendix we provide the Mathematica codes for computing the four-point functions obtained in section \ref{sectionweak}. We implement both the brute force result \eqref{4ptbruteforce} as well as the integrability-based formula \eqref{tail} and give some explicit examples at the end of this appendix. With some slight modifications of the code presented below, equation \eqref{tail2} can be implemented in a similar manner, but we leave that to the interested reader.

\subsection*{Some useful functions}
\begin{flushleft}
{\small
\verb"Le=Length;"\\
\verb"f[u_]=1+I/u; g[u_]=I/u; n[0]=0; m[0]=0;"\\
\verb"f[l1_List,l2_List]:=Product[f[l1[[j1]]-l2[[j2]]],{j1,Le[l1]},{j2,Le[l2]}]"\\
\verb"h[l1_List,l2_List]:=Product[h[l1[[j1]]-l2[[j2]]],{j1,Le[l1]},{j2,Le[l2]}]"\\
\verb"fs[l1_List]:=Product[f[l1[[j1]]-l1[[j2]]],{j1,Le[l1]},{j2,j1+1,Le[l1]}]"\\
\verb"gs[l1_List]:=Product[g[l1[[j1]]-l1[[j2]]],{j1,Le[l1]},{j2,j1+1,Le[l1]}]"\\
\verb"fb[l1_List]:=Product[f[l1[[j1]]-l1[[j2]]],{j1,Le[l1]},{j2,j1-1}]"\\
\verb"gb[l1_List]:=Product[g[l1[[j1]]-l1[[j2]]],{j1,Le[l1]},{j2,j1-1}]"\\
\verb"gp[l_List]:=Times@@g[l+I/2]; gm[l_List]:=Times@@g[l-I/2]"\\
\verb"e[l_List]:=Times@@((l+I/2)/(l-I/2))"\\
\verb"Dvd[ls_List]:=({Complement[ls,#],#}&)/@Subsets[ls,{0,Le[ls]}];"\\
\verb"BPS[N_]:=Table[10^(10+j),{j,N}];"
}
\end{flushleft}

\subsection*{Four-point functions by brute force (\ref{4ptbruteforce})}
\begin{flushleft}
{\small
\verb"Off[Det::matsq];"\\ 
\verb"S[x_,y_]:=(x-y+I)/(x-y-I)"\\
\verb"Wave[l_List]:=Block[{p=Permutations[Range[Le[l]]],i,j},Sum[A[p[[i]]]"\\
\verb"Product[((l[[p[[i,j]]]]+I/2)/(l[[p[[i,j]]]]-I/2))^n[j],{j,1,Le[l]}],{i,1,Le[p]}]"\\
\verb"//.{A[{a___,b_,c_,d___}]:>S[l[[b]],l[[c]]]A[{a,c,b,d}]/;b>c}"\\
\verb"/.{A[a___]:>1/;a==Range[Le[a]]}];"\\ 
\verb"dphi[L_,l_List]:=Det@Table[-If[i==j,L/(l[[i]]^2+1/4)-Sum[2/((l[[i]]-l[[k]])^2+1),"\\
\verb"{k,Le[l]}],0]-2/(1+(l[[i]]-l[[j]])^2),{i,Le[l]},{j,Le[l]}]/.Det[{}]->1"\\
\verb"normdet[L_,l_List]:=Block[{inf,fin,onlyfin},inf=Select[l,Abs[#]>10^8&];"\\ 
\verb"If[inf=={},fin=l,fin=Select[l,MemberQ[inf,#]==False&]];"\\ 
\verb"onlyfin=(fb[fin]dphi[L,fin])/(fb[Conjugate[fin]]gp[fin]gm[fin]);"\\ 
\verb"If[inf=={},onlyfin,(((L-2Le[fin])!Le[inf]!)onlyfin)/(L-2Le[fin]-Le[inf])!]]"\\
\verb""\\
\verb"ClearAll[C1234bf]"\\
\verb"C1234bf[L1_,N1_,L2_,N2_,L3_,N3_,L4_,N4_,l1_List,l2_List,l3_List,l4_List]:="\\ 
\verb"Block[{j,k,psis,norms,lim,lim2},psis[r_,s_]:=(Wave[l1]/.{n[j_]:>r+j-s/;s<j<=s+N4,"\\
\verb"n[k_]:>N4+r+m[k-s-N4]/;s+N4<k<=N1-N3,n[q_]:>L1+q-N1/;q>N1-N3})"\\ 
\verb"(Wave[l2]/.{n[k_]:>L2-(L4-N4+r+m[N1-N3-N4-s-k+1])+1/;k<=N1-(N3+N4+s),"\\ 
\verb"n[j_]:>L2-n[N1-(N3+N4)-j+1]+1/;j>N1-(s+N3+N4)})(Wave[l3]/.n[j_]->j)"\\
\verb"(Wave[l4]/.n[j_]->L4-N4+j);"\\
\verb"norms=normdet[L1,l1]normdet[L2,l2]normdet[L3,l3]normdet[L4,l4];"\\
\verb"lim[s_,r_]:=Sequence@@Table[{n[j],n[j-1]+1,r},{j,s}];"\\
\verb"lim2[s_,r_]:=Sequence@@Table[{m[k],m[k-1]+1,L1-N3-N4-r},{k,N1-N3-N4-s}];"\\
\verb"Sqrt[L1 L2 If[L3==0,1,L3]If[L4==0,1,L4]/norms]"\\
\verb"Sum[Sum[If[L3==0 || L4==0,1/(L1-N3-N4+1),1]psis[r,s],lim2[s,r],lim[s,r]],"\\
\verb"{r,0,L1-N3-N4},{s,0,Min[r,N1-N3-N4]}]]"
}
\end{flushleft}

\subsection*{Four-point functions from integrability (\ref{tail})}
\begin{flushleft}
{\small
\verb"AtoC[u_List,v_List,L_]:=Block[{a,d,f,g,z},a[z_]=(z + I/2)^L;d[z_]=(z-I/2)^L;"\\
\verb"f[z_]=(z + I)/z;g[z_]=I/z;(1/(Times@@g[u+I/2]Times@@g[v-I/2]"\\
\verb"Times@@a[v] Times@@d[u]))(1/(Product[f[u[[j1]]-u[[j2]]],{j1,Le[u]},"\\ 
\verb"{j2,j1+1,Le[u]}]Product[f[v[[j1]]-v[[j2]]],{j1,Le[u]},{j2,j1-1}]))]"\\
\verb"ClearAll[NewrecAl];"\\ 
\verb"NewrecAl[u_,v_,L_]:=NewrecAl[u,v,L]=Module[{N=Le[u],n,i,j,a,d,z,f,g,n1,n2},"\\
\verb"a[z_]=(z+I/2)^L; d[z_]=(z-I/2)^L; f[z_]=(z+I)/z; g[z_]=I/z;"\\ 
\verb"Sum[g[u[[1]]-v[[n]]](Product[If[j!=n,f[u[[1]]-v[[j]]],1],{j,1,N}]"\\
\verb"Product[If[j!=n,f[v[[j]]-v[[n]]],1],{j,1,N}]a[v[[n]]]d[u[[1]]]-"\\ 
\verb"Product[If[j!=n,f[v[[j]]-u[[1]]],1],{j,1,N}]"\\
\verb"Product[If[j!=n,f[v[[n]]-v[[j]]],1],{j,1,N}]a[u[[1]]]d[v[[n]]])"\\
\verb"NewrecAl[Drop[u,1],Drop[v,{n}],L],{n,1,N}]-"\\ 
\verb"If[Le[v]==1,0,Sum[(g[u[[1]]-v[[n1]]]g[u[[1]]-v[[n2]]])"\\
\verb"(Product[f[v[[n1]]-j]f[j-v[[n2]]],{j,Join[Take[v,n1-1],"\\ 
\verb"Take[v,{n1+1,n2-1}],Take[v,{n2+1,Le[v]}]]}]f[v[[n1]] - v[[n2]]]"\\
\verb"a[v[[n2]]]d[v[[n1]]]+Product[f[j-v[[n1]]]f[v[[n2]]-j],"\\ 
\verb"{j,Join[Take[v,n1-1],Take[v,{n1+1,n2-1}],"\\ 
\verb"Take[v,{n2+1,Le[v]}]]}]f[v[[n2]]-v[[n1]]]a[v[[n1]]] d[v[[n2]]])"\\
\verb"NewrecAl[Drop[u,1],Join[{u[[1]]},Join[Take[v,n1-1],Take[v,{n1+1,n2-1}],"\\ 
\verb"Take[v,{n2+1,Le[v]}]]],L],{n2,2,N},{n1,1,n2-1}]]]"\\
\verb"NewrecAl[{}, {}, L_] = 1;"\\
\verb"SProduct[u_List,v_List,L_]:=AtoC[u,v,L]NewrecAl[u,v,L]"\\ 
\verb""\\
\verb"ClearAll[C1234int]"\\
\verb"C1234int[le_,L1_,N1_,L2_,N2_,L3_,N3_,L4_,N4_,l1_List,l2_List,l3_List,l4_List]:="\\ 
\verb"Block[{a1,a1b,a2,a2b,a3,a3b,b1,b1b,b2,b2b,b3,b3b,dv1,dv2,dv3,dv4,norms},"\\ 
\verb"dv1=Dvd[l1]; dv2=Dvd[l2];"\\ 
\verb"norms=normdet[L1, l1]normdet[L2, l2]normdet[L3,l3]normdet[L4,l4];"\\
\verb"Sqrt[(L1 L2 If[L3==0,1,L3]If[L4==0,1,L4])/norms]"\\
\verb"Sum[If[L4==0||L3==0,1/(L1-N3-N4+1),1](a1=dv1[[i,1]];a1b=dv1[[i,2]];"\\
\verb"b2=dv2[[j,1]];b2b=dv2[[j,2]];If[Le[a1]==Le[b2b],dv3=Dvd[a1b];"\\
\verb"Sum[a2=dv3[[k,1]];a2b=dv3[[k,2]];If[Le[a2]==N4,dv4=Dvd[a2b];Sum[a3=dv4[[l,1]];"\\
\verb"a3b=dv4[[l,2]];If[Le[a3b]==N3&&Le[a3]==Le[b2],(1/(fs[l1]fs[l2]))"\\
\verb"e[a1b]^(L1-N3-N4-le)e[a2b]^N4 e[a3b]^(N3+le+1)e[l2]^(L3-N3+le+1)"\\
\verb"e[b2b]^(L2-L3+N3-le)e[l4]^(L4+1)f[a1,a1b]fs[a1]f[a2,a2b]fs[a2]f[a3,a3b]fb[a3b]"\\
\verb"fs[a3]f[b2,b2b]fb[b2]fb[b2b](fb[l4]/fs[l4])SProduct[a3,b2,le]"\\
\verb"SProduct[a1,b2b,L1-N3-N4-le]SProduct[a2,l4,N4]SProduct[l3,a3b,N3],0],"\\
\verb"{l,1,Le[dv4]}],0],{k,1,Le[dv3]}],0]),{i,1,Le[dv1]},{j,1,Le[dv2]}]];"\\
\verb"C1234int[L1_,N1_,L2_,N2_,L3_,N3_,L4_,N4_,l1_List,l2_List,l3_List,l4_List]:="\\
\verb"(temp[le_]=C1234int[le,L1,N1,L2,N2,L3,N3,L4,N4,l1,l2,l3,l4];Sum[temp[le],"\\ 
\verb"{le,0,L1-N3-N4}])" 
}
\end{flushleft}

\subsection*{Examples}
Let us now show how to use the code to compute four-point functions. Note that the functions \verb"C1234bf" and \verb"C1234int" can be used to compute any four-point function obeying the setup of figure \ref{setuppartitions}. In the examples below, \verb"us", \verb"vs", \verb"ws" and \verb"zs" are the Bethe roots of operators $\O_1$, $\O_2$, $\O_3$ and $\O_4$, respectively, for the charges indicated in each case.\footnote{Again, it is computationally easy to find the positions of a very large number of Bethe roots, see section 7 of \cite{Bargheer:2008kj}.} Let us consider first an example in which all four operators are non-BPS. Running the following code
\begin{flushleft}
{\small
\verb"L1=12;N1=6;L2=12;N2=2;L3=4;N3=2;L4=4;N4=2;"\\
\verb"us={0.676245041405-0.993633391204 I,0.678017442247,0.676245041405+0.993633391204 I,"\\ 
\verb"-0.676245041405+0.993633391204 I,-0.678017442247,-0.676245041405-0.993633391204 I};"\\
\verb"vs={1.702843619444,-1.702843619444};"\\
\verb"ws={0.288675134594,-0.288675134594};"\\
\verb"zs={0.288675134594,-0.288675134594};"\\
\verb"C1234bf[L1,N1,L2,N2,L3,N3,L4,N4,us,vs,ws,zs]"\\
\verb"C1234int[L1,N1,L2,N2,L3,N3,L4,N4,us,vs,ws,zs]"\\
}
\end{flushleft}
we obtain a perfect agreement between the brute force computation and the integrability-based formula, giving
\beqq
C^{\bullet \bullet \bullet \bullet}_{1234}=1.28031-0.66373\, i \, .
\eeqq
As we mentioned in the main text of the paper, the integrability-based formula is computationally much faster than the brute force formula. This can be checked in Mathematica for the example above by simply using the function \verb"AbsoluteTiming" when executing \verb"C1234bf" and \verb"C1234int". The result is that the former takes $18$ seconds to compute, while the latter only takes $1$ second!  Of course, the gain in efficiency is even more notorious as we increase the length and number of excitations of each operator in the four-point function.

Now, let us consider a case in which $\O_3$ and $\O_4$ are BPS operators. The following configuration corresponds to the second point of data for $\alpha=1/3$ used in section \ref{sectionmatch}. Running
\begin{flushleft}
{\small
\verb"L1=18;N1=6;L2=18;N2=2;L3=4;N3=2;L4=4;N4=2;"\\
\verb"us={1.892118528217-1.114121041954 I,1.976939483361,1.892118528217+1.114121041954 I,"\\ 
\verb"-1.892118528217+1.114121041954 I,-1.976939483361,-1.892118528217-1.114121041954 I};"\\
\verb"vs={2.674763752754,-2.674763752754};"\\
\verb"ws={10^10,10^15};"\\
\verb"zs={10^11,10^16};"\\
\verb"C1234int[L1,N1,L2,N2,L3,N3,L4,N4,us,vs,ws,zs]"\\
}
\end{flushleft}
we obtain
\beqq
C^{\bullet \bullet \circ \circ}_{1234}=-5.38224-6.23581\, i \, .
\eeqq
Upon diving this result by the product $C^{\circ \circ \circ}_{123} C^{\circ \circ  \circ}_{124}$, which can be implemented using the code presented in appendix D of \cite{paper1}, and taking the absolute value of the resulting ratio, we get $r_{1234}=0.65401$, which is exactly the second blue bullet plotted in figure \ref{dataplot}. 

Finally, we should mention that when some of the operators in the four-point function are BPS, it is computationally more efficient to replace the appropriate general scalar products \verb"SProduct" (which compute formula \eqref{newrecco}) appearing in \verb"C1234int" by a function that implements the scalar product between a Bethe state and a BPS state \eqref{scalarBPS}.


\end{document}